\pdfoutput=1
\documentclass{egpubl}
\usepackage{pg2023}

\SpecialIssuePaper         %

\CGFStandardLicense

\usepackage[T1]{fontenc}
\usepackage{dfadobe}  

\usepackage{cite}  %
\BibtexOrBiblatex
\electronicVersion
\PrintedOrElectronic
\ifpdf \usepackage[pdftex]{graphicx} \pdfcompresslevel=9
\else \usepackage[dvips]{graphicx} \fi

\usepackage[ruled,vlined]{algorithm2e}
\usepackage{amsfonts}       %
\usepackage{amsmath}
\usepackage{amssymb}
\usepackage{bm}
\usepackage{booktabs}       %
\usepackage{breqn}
\usepackage[font=small]{caption}
\usepackage{epsfig}
\usepackage[T1]{fontenc}    %
\usepackage{gensymb}
\usepackage[utf8]{inputenc} %
\usepackage{longtable}
\usepackage{makecell}
\usepackage{microtype}      %
\usepackage{multirow}
\usepackage{nicefrac}       %
\usepackage[group-separator={,}]{siunitx}
\usepackage{tabularx}
\usepackage{times}
\usepackage{url}            %
\usepackage{wrapfig}
\usepackage{xcolor}         %
\usepackage{xspace}
\usepackage[normalem]{ulem} %
\usepackage[outdir=./]{epstopdf} 

\usepackage{filecontents,catchfile,pgffor}
\usepackage{egweblnk}

\newcommand*{\OptCtrlPoints}[0]{\textsc{OptCtrlPoints}}

\SetAlCapHSkip{0pt}
\SetKwInput{KwInput}{Inputs}
\SetKwInput{KwOutput}{Outputs}
\SetKwInOut{KwParam}{Parameters}

\newcolumntype{Y}{>{\centering\arraybackslash}X}

\newcommand{\bA}{\mathbf{A}}
\newcommand{\bB}{\mathbf{B}}
\newcommand{\bC}{\mathbf{C}}
\newcommand{\bD}{\mathbf{D}}

\newcommand{\bI}{\mathbf{I}}

\newcommand{\bM}{\mathbf{M}}

\newcommand{\bP}{\mathbf{P}}
\newcommand{\bQ}{\mathbf{Q}}

\newcommand{\bs}{\mathbf{s}}\newcommand{\bS}{\mathbf{S}}
\newcommand{\bT}{\mathbf{T}}

\newcommand{\bV}{\mathbf{V}}
\newcommand{\bW}{\mathbf{W}}
\newcommand{\bX}{\mathbf{X}}

\newcommand{\bZ}{\mathbf{Z}}

\DeclareMathOperator*{\argmin}{argmin~}

\makeatletter
\DeclareRobustCommand\onedot{\futurelet\@let@token\@onedot}
\def\@onedot{\ifx\@let@token.\else.\null\fi\xspace}
\def\eg{e.g\onedot} 
\def\ie{i.e\onedot}

\def\etal{et~al\onedot}

\makeatother

\renewcommand{\eqref}[1]{Eq.~\ref{#1}}

\definecolor{darkred}{rgb}{0.7,0.2,0.1}
\definecolor{darkgreen}{rgb}{0,0.7,0}
\definecolor{orange}{RGB}{255,127,0}
\definecolor{ourpurple}{RGB}{127,127,204}
\definecolor{palgreen}{RGB}{51,179,179}
\definecolor{magenta}{RGB}{199,21,133}

\definecolor{revcolor}{RGB}{255,0,0}

\definecolor{mkcolor}{RGB}{255,0,128}

\definecolor{khcolor}{RGB}{0,128,128}

\definecolor{despicolor}{RGB}{128, 0, 0}

\definecolor{aleccolor}{RGB}{128,0,128}

\definecolor{leocolor}{RGB}{0,0,255}

\definecolor{mhcolor}{RGB}{0,128,0}

\title{OptCtrlPoints: Finding the Optimal Control Points for Biharmonic 3D Shape Deformation \vspace{-0.6cm}}

\author[Kunho Kim .et al]{Kunho Kim\textsuperscript{*1}\orcid{0009-0006-4087-2577}  $\quad$
Mikaela Angelina Uy\textsuperscript{*2}\orcid{0009-0009-4917-7724}$\quad$
Despoina Paschalidou\textsuperscript{2}\orcid{0009-0002-9079-3835} $\quad$
Alec Jacobson\textsuperscript{3}\orcid{0000-0003-4603-7143} $\quad$
Leonidas J. Guibas\textsuperscript{2}\orcid{0000-0002-8315-4886} $\quad$
Minhyuk Sung\textsuperscript{1}\orcid{0000-0001-7428-9570} \\
\textsuperscript{1}KAIST $\quad$ \textsuperscript{2}Stanford University $\quad$  \textsuperscript{3}University of Toronto}

\volume{42}   %
\issue{7}     %
\pStartPage{1}      %

\begin{document}

\twocolumn[{%
\renewcommand\twocolumn[1][]{#1}%

\maketitle

\begin{center}
    \captionsetup{type=figure, labelfont=bf, textfont=it}
    \vspace{-2\baselineskip}
    \includegraphics[width=\linewidth]{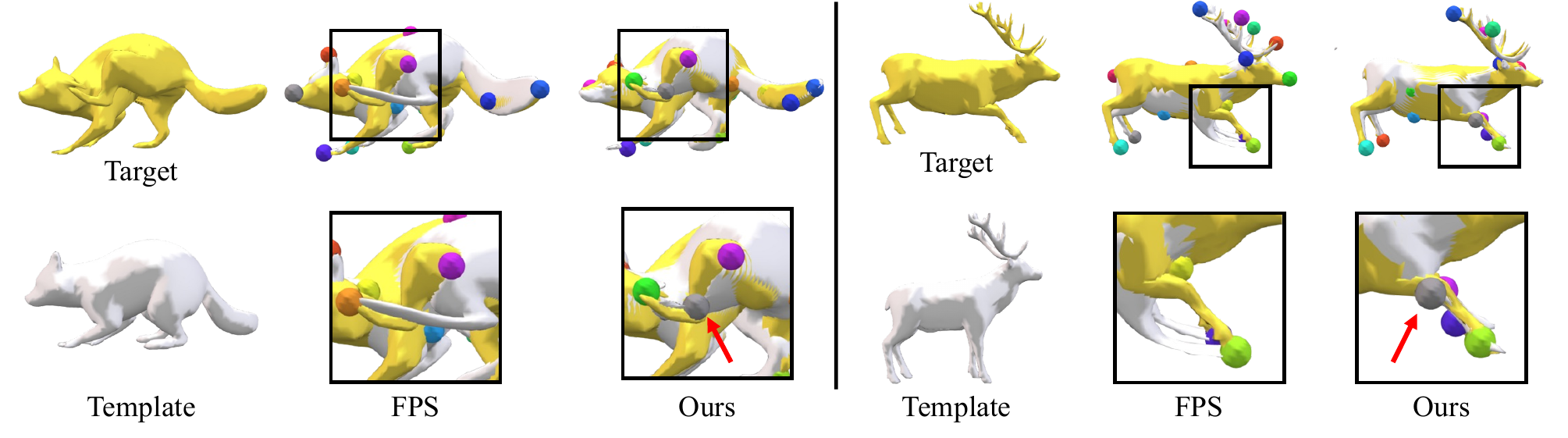}
    \caption{
    \OptCtrlPoints{}, a data-driven method to search for the optimal set of control points, enables accurate replication of the possible variations of a shape through biharmonic deformation. In contrast to control points obtained through Farthest Point Sampling, the control points discovered by \OptCtrlPoints{} are strategically positioned, such as at the knees (see the red arrows in the black boxes), resulting in a superior fit of the template to the targets during deformation.}
    \label{fig:teaser}
    \end{center}
    \vspace{2\baselineskip}
}]

\begin{abstract}

We propose~\OptCtrlPoints{}, a data-driven framework designed to identify the optimal sparse set of control points for reproducing target shapes using biharmonic 3D shape deformation. Control-point-based 3D deformation methods are widely utilized for interactive shape editing, and their usability is enhanced when the control points are sparse yet strategically distributed across the shape. With this objective in mind, we introduce a data-driven approach that can determine the most suitable set of control points, assuming that we have a given set of possible shape variations.
The challenges associated with this task primarily stem from the computationally demanding nature of the problem. Two main factors contribute to this complexity: solving a large linear system for the biharmonic weight computation and addressing the combinatorial problem of finding the optimal subset of mesh vertices. To overcome these challenges, we propose a reformulation of the biharmonic computation that reduces the matrix size, making it dependent on the number of control points rather than the number of vertices. Additionally, we present an efficient search algorithm that significantly reduces the time complexity while still delivering a nearly optimal solution.
Experiments on SMPL, SMAL, and DeformingThings4D datasets demonstrate the efficacy of our method. Our control points achieve better template-to-target fit than FPS, random search, and neural-network-based prediction. We also highlight the significant reduction in computation time from days to approximately 3 minutes.


\end{abstract}

\section{Introduction}
\label{sec:intro}
\footnotetext{\textsuperscript{*} denotes equal contribution.}

The demand for high-quality 3D models is growing rapidly, especially with recently emerging applications in virtual and augmented realities, gaming, robotics, animation, etc. However, creating and designing high-quality 3D models is a tedious and difficult process even for expert designers. 
Shape deformation is thus an important technique that enables producing plausible variations of existing high-quality, artist-generated 3D assets.

Deforming a 3D model is, however, a highly non-trivial task. A straightforward approach is to parameterize deformation as positions of all the mesh vertices~\cite{CycleConsistency,Sung:2020,3dn}, although it is difficult for users to edit the shape and also can lead to unrealistic outputs due to its large degree of freedom. To circumvent this conundrum, existing works in geometry processing~\cite{biharmonic,Wang:2015:Linear,green,Ju_meanvalue,Joshi:2007,Hormann2008MaximumEC,webber2011,li2013,weber2007context,baran2007automatic} leverage a \emph{sparse} set of \emph{deformation handles} to constrain and parameterize deformation within a lower degree of freedom, facilitating more intuitive editing via interactions with the users.

For handle-based deformation to be effective and useful, there are several desirable properties, such as identity (\ie preserving the shape under zero handle movement), locality, smoothness, closed-form expression, and flexibility for the representation of shapes. Due to such desirable properties, many existing deformation methods~\cite{biharmonic,Wang:2015:Linear} use a set of \emph{points} or \emph{regions} in the mesh, which is typically a subset of the mesh vertices, as the handles (Fig.~\ref{fig:teaser}) while defining the shape deformation function from the handles using \emph{biharmonic} weights. Given any input shape, point and region handles can be directly selected, and their biharmonic weights can also be directly computed, offering flexibility and convenience. This is in contrast to other types of deformation handles, such as cages~\cite{green,Ju_meanvalue,Joshi:2007,Hormann2008MaximumEC,webber2011,li2013} and skeletons~\cite{weber2007context,baran2007automatic}. Cages require the manual construction of a closed polyhedral envelope for the shapes, while skeletons require rigging, where the skeleton structure needs to be delicately constructed to produce detailed deformation and typically necessitate manual weight painting.

Given a mesh and a sparse subset of the mesh vertices representing the control points, the biharmonic weights defining the linear map from the control point positions to the mesh vertex positions are calculated by solving a convex quadratic optimization problem~\cite{biharmonic}, which was shown to be equivalent to solving a linear system~\cite{Wang:2015:Linear} when relaxing some constraints.
To obtain a wide range of plausible variations of the shape from a sparse set of control points, it is crucial to find the \emph{optimal} set of control points. For an articulated 3D shape, for instance, the control points near the joints would be able to produce more realistic deformations by bending the joints properly (see control points at knees on the left of Fig.~\ref{fig:teaser}). Also, more control points would be needed in the regions with more detailed variations (see the additional control points on the leg on the right of Fig.~\ref{fig:teaser}).

In this work, we present a \emph{data-driven} method for finding the optimal set of control points, coined~\OptCtrlPoints{}. Given a template mesh and its variant shapes (\eg different poses in an animation), we find the ideal subset of the template mesh vertices as control points that can best fit the template to all the variant shapes via deformation. Our method can thus provide optimized deformation handles to the users and enables easier shape editing. In contrast to previous works~\cite{Yifan:NeuralCage:2020,jakab2021keypointdeformer} that utilize neural networks to learn deformation handles in a data-driven manner, our approach focuses on discovering a set of control points rather than fitting a sphere cage to the template mesh. The limitation of using a sphere cage is that it is unable to accommodate large deformations, making it unsuitable for non-rigid shapes like human or animal bodies. By identifying control points instead, our method enables more flexible and effective deformation modeling for such shapes.

Finding the optimal set of vertices poses a challenge due to the substantial amount of computation time involved, which can span several hours or more than a day. Two primary factors contribute to this extended computation duration.
Firstly, the process of deforming the template mesh to accurately align with the target shapes is time-consuming. While deforming the mesh using \emph{fixed} biharmonic weights can be performed quickly by prefactorizing a matrix within a linear system, the need to test different sets of control points prevents prefactorization of the matrix, leading to a considerably slower process.
Secondly, solving a combinatorial optimization problem to determine the ideal subset of vertices becomes intractable when dealing with thousands or more vertices. When $N$ is the number of vertices and $K$ is the number of control points, a straightforward exhaustive search requires a time complexity of $\mathcal{O}(N^K)$,  further contributing to the computational challenges.

To overcome these challenges and address the issue of intractable computation, we propose a novel algorithm that incorporates two key ideas. Firstly, we introduce a reformulation of the biharmonic weight computation, which significantly reduces the time required to solve the linear system. This is achieved by introducing a new linear system where the size of the matrix is not dependent on the number of vertices $N$, but rather on the number of control points $K$. This reformulation proves particularly effective in cases where the set of control points varies, as in our scenario where we search for the best set.
Additionally, we present an efficient search algorithm that leverages the new biharmonic weight computation. This algorithm operates by iteratively updating control points one by one while simultaneously traversing local partitioned regions for each control point. This simple yet effective approach enables us to reduce the time complexity from $\mathcal{O}(N^K)$ to $\Theta(N+K^2)$, which is linear order of the number of the vertices while still providing a nearly optimal solution.

In our experiments, we assess the performance of our method by evaluating the alignment of the template to the target through deformation, comparing it with other baselines: Farthest Point Sampling (FPS), random search, and KeypointDeformer~\cite{jakab2021keypointdeformer}, a neural-network-based method for predicting keypoints for deformation. We conduct these evaluations on three datasets: SMPL~\cite{SMPL:2015}, SMAL~\cite{Zuffi:CVPR:2017}, and DeformingThings4D~\cite{deformthings4d}.
Our results demonstrate that the control points discovered by our \OptCtrlPoints{} algorithm offer a better fit of the template to the target shapes, thanks to their ideal locations for producing the desired variations. Additionally, our approach significantly reduces computation time, especially when compared to random search. In approximately 3 minutes, our method can find a good set of control points, whereas without the new biharmonic weight formulation and efficient search, it would take days to achieve a similar outcome.
Furthermore, with the DeformingThings4D~\cite{deformthings4d} dataset, we illustrate that our data-driven control point search method can discover an optimized set of control points tailored to the given set of target shapes. We conduct experiments using two setups: targeting all motions or focusing on a specific motion within the animations. The consistent lower fitting errors observed in the specific motion case compared to the all motion case highlight the effectiveness of our data-driven approach.

\section{Related work}
\label{sec:related}

\subsection{3D Shape Deformation} 3D shape deformation has been a long-standing problem in computer graphics and geometry processing. The problem is to find the best vertex positions for a given mesh in order to obtain a new shape that best fits a target while preserving the local geometric details of the original shape. Previous approaches include free-form deformation~\cite{Kraevoy:2008, Sederberg:1986} that define a smooth deformation function by interpolating the weights of the voxel grids enclosing the surface, and vertex-based approaches~\cite{Sorkine:2004,Igarashi:2005,Sorkine:2007, Lipman:2004,Lipman:2005} where vertex positions are directly optimized through a target-fitting objective function. Regularization losses are also used, such as mesh Laplacian~\cite{Sorkine:2004,Lipman:2004} and local rigidity~\cite{Lipman:2005,Igarashi:2005,Sorkine:2007} to preserve the geometric details of the original shape. Learning-based approaches have also been introduced for both free-form~\cite{Yumer:2016,Hanocka:2018,Jack:2018,Kurenkov:2018} and vertex-based~\cite{CycleConsistency, 3dn} deformation approaches. More recently, neural implicit functions~\cite{deng2021deformed, zheng2021deep} explore defining the deformation offset on the full coordinate space, instead of only on the surface of the mesh. These works, however, may lead to unrealistic outputs due to their large degrees of freedom and also
do not exert intuitive control over the shape, as editing operations are performed in the implicit space.

\subsection{Traditional Handle-based Shape Deformation} Deformation handles are commonly used to address the need for low-dimensional control on shape deformation and have been well studied in the computer graphics literature~\cite{sorkine09}. Earlier works use volumetric prisms~\cite{botsch06, botsch08} or off-surface handles~\cite{bostch2007} to compute detail-preserving shape deformation through variational methods.
These variational methods typically require optimization at each time of deformation.
Cage~\cite{Ju_meanvalue, Joshi:2007, Hormann2008MaximumEC, webber2011, li2013, green} is another form of shape handles where a shape is enclosed in a coarse polytope, and the mesh vertices are defined as a linear combination of the cage vertices through generalized barycentric coordinates.
Skeletons~\cite{weber2007context, baran2007automatic} also define a linear map from the joints and bones to the mesh vertices via linear blend skinning, while now the handles appear inside the shape.  
Both cages and skeletons allow for shape deformation to be expressible in a closed form, but require manual construction of the source cage or rigging. Jacobson \etal~\cite{biharmonic} introduced a handle-based deformation function based on solving the \emph{biharmonic equation} over the mesh surface with boundary constraints that can use a set of points or regions in the mesh as handles to define shape deformation using biharmonic weights. Unlike cages and skeletons, these handles can directly be computed for any source shape and are thus flexible and versatile. Wang \etal~\cite{Wang:2015:Linear} then introduced a closed-form formulation to the original constrained quadratic optimization formulation.
In our method, we leverage this versatile deformation handle with an efficient reformulation to enable tractable gradient-based optimization for handle \emph{discovery}, in contrast to existing works that assume shape handles to be given. 

\subsection{Learning Handle-based Shape Deformation} Recently, handle-based shape deformations have been revisited in the context of deep learning. DeformSyncNet~\cite{Sung:2020} uses rigid bounding boxes as shape handles for learning a latent shape difference deformation space.  Wang \etal~\cite{Yifan:NeuralCage:2020} introduced Neural Cages, a neural network for cage-based deformations that predicts a source-dependent cage used to deform a source shape to match a given target. KeypointDeformer~\cite{jakab2021keypointdeformer} leverages Neural Cages to learn keypoints that can be used for deformation. However, the degree of plausible output shape deformations yielded by the neural cage-based deformation methods is limited since they start from a sphere-based cage, making highly non-rigid deformations on non-sphere-like shapes difficult. In contrast, we leverage a deformation function defined with biharmonic coordinates that enable flexibility for any given source shape. Liu \etal \cite{DeepMetaHandles:2021} also use biharmonic coordinates as
their deformation function. In contrast to our work, where the goal is the \emph{discovery} of the control points, they assume them to be given and instead learn a latent space of \emph{meta-handles} for the given set of control points. Moreover, we discover \emph{explicit} handles, which are 3D mesh vertices, that allow direct user interpretability and controllability, instead of meta-handles in latent space.

\section{Background}
\label{sec:method}

\subsection{Shape Deformation and Deformation Handles}
\label{sec:deformation_handles}
The creation of high-quality 3D models is a tedious process that requires manual expertise, thus making shape deformation an important task as it enables converting an existing 3D model to a new shape while preserving their fine details. However, naive mesh deformation through moving individual mesh vertices is cumbersome as it is both difficult for users and can easily lead to unrealistic outputs. Existing work~\cite{biharmonic,Wang:2015:Linear,green,Ju_meanvalue,Joshi:2007,Hormann2008MaximumEC,webber2011,li2013,weber2007context,baran2007automatic} thus introduce intuitive \emph{deformation handles}, \eg control points, cages, skeletons, etc., to constrain and parameterize deformation with a low degree of freedom. Several properties are critical for handle-based deformation to be effective and useful:
\begin{enumerate}
    \item \textbf{Identity}: The original shape must be reconstructed under zero movement of shape handles.
    \item \textbf{Locality}: The deformation produced by each individual handle must be local and smooth.
    \item \textbf{Closed-form}: The output deformed shape must be expressed in a closed-form given the transformations of the deformation handles.
    \item \textbf{Flexibility}: The deformation handles and function must be defined without any constraints or additional information about the shape (\eg, a cage, a skeleton, or bounding primitives).
\end{enumerate}

For these reasons, many existing shape deformation methods~\cite{DeepMetaHandles:2021,lee2022popoutmotion,biharmonic,Wang:2015:Linear} use biharmonic weights as the deformation function. Jacobson \etal~\cite{biharmonic} first introduced bounded biharmonic coordinates that solve biharmonic equations defined over a mesh
to compute the linear map from the handles to the mesh vertices. Wang \etal~\cite{Wang:2015:Linear} later introduced a closed-form formulation for the biharmonic coordinate-based deformation, which we base our work on.
Below, we explain the details about the closed-form formulation of the biharmonic coordinate deformation function.

\begin{figure*}[!t]
\captionsetup{type=figure, labelfont=bf, textfont=it}
\includegraphics[width=\linewidth]{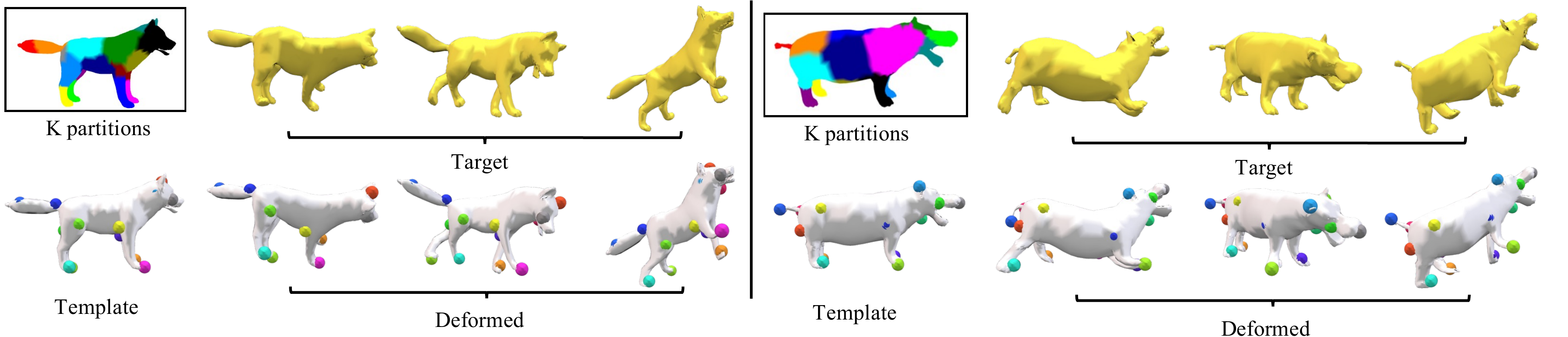}
\caption{
Examples of the template and target shapes, along with the fitting results obtained through biharmonic deformations from the template to the target. The colored points indicate the control points used for the deformation computation. The segmented shapes within the black boxes at the top left corner illustrates the partitioned volume of the source tetrahedral mesh, enabling a level-of-detail search for the optimal placement of each control point.
}
\label{fig:pipeline}
\end{figure*}

\subsection{Biharmonic Coordinates}
\label{sec:orig_biharmonic_coor}
Given a 3D volumetric mesh with $N$ vertices and $K$ control points, which is a sparse \emph{subset} of the mesh vertices ($K \ll N$) represented with a binary selector matrix $\bS \in \{0, 1\}^{K \times N}$,
the biharmonic deformation function~\cite{biharmonic, Wang:2015:Linear} from the positions of the control points $\bC \in \mathbb{R}^{K \times 3}$ to the positions of mesh vertices $\bV \in \mathbb{R}^{N\times 3}$ is defined as a \emph{linear} function: $\bV=\bW\bC$. (The original biharmonic deformation~\cite{biharmonic, Wang:2015:Linear} supports region handles, while we limit our scope to point handles for simplicity.)
Here, $\bW \in \mathbb{R}^{N\times K}$ called \emph{biharmonic weights} is derived from the solution of the following optimization for the vertex positions $\bV$ with respect to the equality constraints on the control point positions:%
\begin{equation}
    \bV = \underset{\bX \in \mathbb{R}^{N \times 3}}{\text{argmin}}
        \frac{1}{2} \operatorname{trace}\left(\bX^{\top} \mathbf{A X}\right) \text { subject to } \mathbf{S X}=\mathbf{C},
    \label{eq:biharmonics_optimization_objective}
\end{equation}
\noindent where $\bA \in \mathbb{R}^{N \times N}$ denotes the discrete Bilaplacian matrix of the mesh.
This optimization finds the positions of the mesh vertices $\bV$ minimizing squared Laplacian energy when the positions of the selected control points are fixed to be $\bC$.
Since the Bilaplacian matrix $\bA$ is positive semi-definite, the optimization is a convex quadratic programming problem that can be solved as a linear system as described in \cite{Wang:2015:Linear}.
The solution thus has a form of $\bV=\bW\bC$ where
\begin{equation}
    \bW (\bS; \bA) = \mathbf{S}^{\top}-\mathbf{T}^{\top}\left(\mathbf{T A T}^{\top}\right)^{-1} \mathbf{T A S ^ { \top }},
    \label{eq:biharmonics_solution}
\end{equation}
and $\bT
\in \{0, 1\}^{(N-K) \times N}$ is the complementary selector matrix of $\bS$ indicating the mesh vertices that are \emph{not} selected as control
points.
By the definition of the Bilaplacian matrix $\bA$ including the original positions of the vertices $\overline{\bV}$ in its null space, the given pose of the mesh $\overline{\bV}$ becomes the solution of the optimization when the control points are not moved ($\bC = \bS\overline{\bV}$), satisfying the identity condition. 

To leverage the expressivity of biharmonic coordinates for shape deformation, we need to find the \emph{optimal} set of control points (a sparse subset of the mesh vertices) that allows us to achieve a wide variety of plausible variants. In what follows, we introduce our \emph{data-driven} approach of finding the optimal set of control points efficiently given possible variants of the shape.

\section{OptCtrlPoints}

\subsection{Problem Definition}
\label{sec:problem_definition}

Given a template volumetric mesh with $N$ vertices which Bilaplacian matrix is $\bA \in \mathbb{R}^{N\times N}$, and a set of $M$ target shapes $\{\mathcal{X}_i\}_{i=1}^{M}$ that are the possible variants of the template, our goal is to find the optimal $K$-subset of the $N$ source vertices as the control points, denoted as $\tilde{\bS} \in \{0, 1\}^{K \times N}$, which can best fit all the target shapes with their corresponding positions in the target (see Fig.~\ref{fig:pipeline}):
\begin{align}
\label{eq:loss}
    {\tilde{\bS}} = \argmin_{\bS \in \{0, 1\}^{K \times N}} & \sum_{i=1}^{M} d(\mathcal{X}_i, \bW(\bS; \bA) \bC(\bS; \mathcal{X}_i)) \nonumber\\
    \text{s.t.}&\quad \sum_{j=1}^{N} \bS_{ij} = 1  \quad \text{for all} \,\, i,
\end{align}
\noindent where $d(\cdot, \cdot)$ is a shape-to-shape distance function, $\bW(\bS; \bA)$ is the biharmonic weight function in Eq.~\ref{eq:biharmonics_solution} for the control point selector matrix $\bS$ and the given Bilaplacian matrix $\bA$, 
and $\bC(\bS; \mathcal{X}_i)$ is a function computing the corresponding positions of the control points $\bS$ in the target $\mathcal{X}_i$.
Assuming that the point-wise map from each target shape $\mathcal{X}_i$ to the template is given or estimated with an off-the-shelf shape correspondence method (\eg, functional maps~\cite{functionalmap}, or its neural variants~\cite{litany2017deep, halimi2019unsupervised, donati2020deep, zeng2021corrnet3d, attaiki2021dpfm, magnet2022smooth}; see a survey~\cite{shape_corr_survey}), let $f: \{\mathcal{X}_i\}_{i=1}^{M} \rightarrow \mathbb{R}^{N\times 3}$ denote a function returning corresponding points of each vertex of the template in the same order. Then $\bC(\bS; \mathcal{X}_i) = \bS f(\mathcal{X}_i)$.

The main challenge in finding the optimal $K$-subset of template vertices as control points in Eq.~\ref{eq:loss} lies in the extensive computation time required.
Note that when the control points and resulting biharmonic weights are \emph{fixed}, the deformations can be computed quickly by prefactorizing $\mathbf{T A T^{\top}}$ in Eq.~\ref{eq:biharmonics_solution}. However, when computing deformations with \emph{different} sets of control points, it is not feasible to leverage the prefactorization since the complementary selector matrix $\mathbf{T}$ varies. Consequently, computing the biharmonic weight matrix $\mathbf{W}$ in Eq.~\ref{eq:biharmonics_solution} becomes the bottleneck, taking several seconds to compute even once (for detailed analysis, refer to Sec.~\ref{sec:results}). Moreover, identifying the optimal $K$-subset from a pool of $N$ elements is an NP-complete combinatorial optimization problem, making exhaustive search impractical due to the typically high number of vertices, often in the thousands.

To address this challenge, we present our efficient control point search framework, called~\OptCtrlPoints{}. First, in Sec.~\ref{sec:our_reformulation}, we introduce a reformulation of Eq.~\ref{eq:biharmonics_solution} that yields the same biharmonic weight matrix $\mathbf{W}$ while solving a linear system on a much smaller scale, significantly reducing computation time. Second, in Sec.~\ref{sec:search}, we propose an efficient search algorithm that reduces the time complexity of the search from $\mathcal{O}(N^K)$ (for exhaustive search) to $\Theta(N + K^2)$ in average, while effectively finding nearly optimal solutions in practice.

\subsection{Reformulation of $\bW(\bS; \bA)$ (Eq.~\ref{eq:biharmonics_solution})}
\label{sec:our_reformulation}
Let $\bM$ denote the linear system in $\bW(\bS; \bA)$ (Eq.~\ref{eq:biharmonics_solution}): 
\begin{equation}
    \bM = \left(\bT \bA \bT^{\top}\right)^{-1} \bT \bA \bS^{\top}.
    \label{eq:tattas}
\end{equation}

We begin by introducing a reformulation of $\bM$ for the case when $\bA$ is not a Bilaplacian matrix but rather an arbitrary invertible square matrix. However, in our specific scenario, $\bA$ is the Bilaplacian matrix, which is non-invertible and positive-semidefinite. In Section \ref{sec:singularity}, we elaborate on how we address the singularity of $\bA$ in this new formulation.

We choose a permutation matrix $\bP \in \mathbb{R}^{N \times N}$ such that the product of $\bP$
with $\bS$ and $\bT$ becomes $\bS\bP=\left[\mathbf{0}_{K \times (N-K)} \mid
\bI_{K \times K}\right]$ and $\bT\bP=\left[\bI_{(N-K) \times(N-K)} \mid \mathbf{0}_{(N-K)
\times K}\right]$, respectively. Moreover, using $\bP$, we define $\bB=\bP^{\top}\bA\bP$ and 
its inverse $\bD=\bP^{\top} \bA^{-1} \bP$ as follows:
\begin{equation}
    \bB=\left[\begin{array}{ll}
    \bB_{11} & \bB_{12} \\
    \bB_{12}^{\top} & \bB_{22}
    \end{array}\right], \bD=\left[\begin{array}{ll}
    \bD_{11} & \bD_{12} \\
    \bD_{12}^{\top} & \bD_{22}
    \end{array}\right],
    \label{eq:permuted_formulas}
\end{equation}
where $\bB_{11}, \bD_{11} \in \mathbb{R}^{(N-K) \times(N-K)}$, $\bB_{12}, \bD_{12} \in
\mathbb{R}^{(N-K) \times K}$, and $\bB_{22}, \bD_{22} \in \mathbb{R}^{K \times
K}$ are block matrices. Now taking into account that a permutation matrix is an orthogonal matrix (\ie
$\bP^{-1}=\bP^{\top}$) and $\bP\bP^{\top}=\bI$, we can rewrite \eqref{eq:tattas} as
follows:
\begin{equation}
    \begin{aligned}
        \bM &= \left(\bT \bA \bT^{\top}\right)^{-1} \bT \bA \bS^{\top} \\
        &= \left((\bT \bP)\left(\bP^{\top} \bA \bP\right)(\bT \bP)^{\top}\right)^{-1}(\bT \bP)\left(\bP^{\top} \bA \bP\right)(\bS \bP)^{\top} \\
        &=\left((\bT \bP) \bB(\bT \bP)^{\top}\right)^{-1}(\bT \bP) \bB(\bS \bP)^{\top} \\
          &=\left(\left[\begin{array}{lll}
          \bI & \mathbf{0}
          \end{array}\right]\left
          [\begin{array}{ll}
          \bB_{11} & \bB_{12} \\
          \bB_{12}^{\top} & \bB_{22}
          \end{array}\right]\left
          [\begin{array}{l}
          \bI \\
          \mathbf{0}
          \end{array}\right]\right)^{-1}\\
          &\left[\begin{array}{ll}
          \bI & \mathbf{0}
          \end{array}\right]\left 
          [\begin{array}{ll}
          \bB_{11} & \bB_{12} \\
          \bB_{12}^{\top} & \bB_{22}
          \end{array}\right]\left[\begin{array}{l}
          \mathbf{0} \\
          \bI
          \end{array}\right] \\
                &=\bB_{11}^{-1} \bB_{12} .
    \end{aligned}
    \label{eq:ptattas}
\end{equation}

\noindent By using the Schur complement~\cite{haynsworth1968schur}, we can express $\bD$, the inverse of $\bB$ as follows:
\begin{equation}
    \begin{aligned}
        \bD^{-1}=\left[\begin{array}{l|l}\bD_{11} & \bD_{12} \\\hline \bD_{12}^{\top} & \bD_{22}\end{array}\right]^{-1}
        =\left[\begin{array}{l|l}\bB_{11} & \bB_{12} \\\hline \bB_{12}^{\top} & \bB_{22}\end{array}\right],
    \end{aligned}
\end{equation}
where
\begin{equation}
    \begin{aligned}
        \bB_{11} &= (\bD_{11}-\bD_{12} \bD_{22}^{-1} \bD_{12}^{\top})^{-1}, \\
        \bB_{12} &= -\left(\bD_{11}-\bD_{12} \bD_{22}^{-1} \bD_{12}^{\top}\right)^{-1} \bD_{12} \bD_{22}^{-1}, \,\, \text{and}\\
        \bB_{22} &= \bD_{22}^{-1} + \bD_{22}^{-1}\bD_{12}^{\top}\left(\bD_{11}-\bD_{12} \bD_{22}^{-1} \bD_{12}^{\top}\right)^{-1} \bD_{12} \bD_{22}^{-1}.
    \end{aligned}
\end{equation}
\noindent Thus, setting $\bQ=\bD_{11}-\bD_{12} \bD_{22}^{-1} \bD_{12}^{\top}$, we have $\bB_{11}^{-1}=\bQ$ and $\bB_{12}=-\bQ^{-1} \bD_{12} \bD_{22}^{-1}$. Hence we get
\begin{equation}
\label{eq:shur}
    \begin{aligned}
    \bM &=\bB_{11}^{-1} \bB_{12} =-\bD_{12} \bD_{22}^{-1} .
    \end{aligned}
\end{equation}

\noindent Finally, by considering the orthogonality of the permutation matrix $\bP$ and
that $\bP^\top = \begin{bmatrix} \bT \\ \bS \end{bmatrix}$ based on its definition,
we can show that $\bD_{12}=\bT \bA^{-1} \bS^{\top}$ and that $\bD_{22}=\bS
\bA^{-1} \bS^{\top}$ as follows:

\begin{equation}
    \begin{aligned}
    & \bD=\bP^{\top} \bA^{-1} \bP \\
    & =\left[\frac{\bT}{\bS}\right] \bA^{-1}\left[\bT^{\top} \mid \bS^{\top}\right] \\
    & =\left[\begin{array}{l|l}\bT \bA^{-1} \bT^{\top} & \bT \bA^{-1} \bS^{\top} \\\hline \bS \bA^{-1} \bT^{\top} & \bS \bA^{-1} \bS^{\top}\end{array}\right] \\
    & =\left[\begin{array}{c|c}\bD_{11} & \bD_{12} \\\hline \bD_{12}^{\top} & \bD_{22} \cdot\end{array}\right], \\
    \end{aligned}
\end{equation}

\noindent Then, \eqref{eq:shur} becomes:
\begin{equation}
    \bM = -\bT \bA^{-1} \bS^{\top}\left(\bS \bA^{-1} \bS^{\top}\right)^{-1}.
    \label{eq:final_m}
\end{equation}

\noindent By replacing \eqref{eq:final_m} in our initial expression from
\eqref{eq:biharmonics_solution}, we obtain the following reformulation:
\begin{equation}
    \bW (\bS; \bA) = \bS^{\top} + \bT^{\top}\left(\bT \bA^{-1} \bS^{\top}\left(\bS \bA^{-1} \bS^{\top}\right)^{-1}\right).
    \label{eq:our_reformulation}
\end{equation}

\noindent Note that Eq.~\ref{eq:our_reformulation} includes a linear system with a significantly smaller matrix, $\bS\bA^{-1}\bS^{\top} \in \mathbb{R}^{K \times K}$, where $K \ll N$. Also, $\bA^{-1}$ can be precomputed to speed up the computation at each iteration.

\subsubsection{Handling the Singularity of the Bilaplacian Matrix $\bA$}
\label{sec:singularity}
\begin{figure}[!t]
\captionsetup{type=figure, labelfont=bf, textfont=it}
\includegraphics[width=\linewidth]{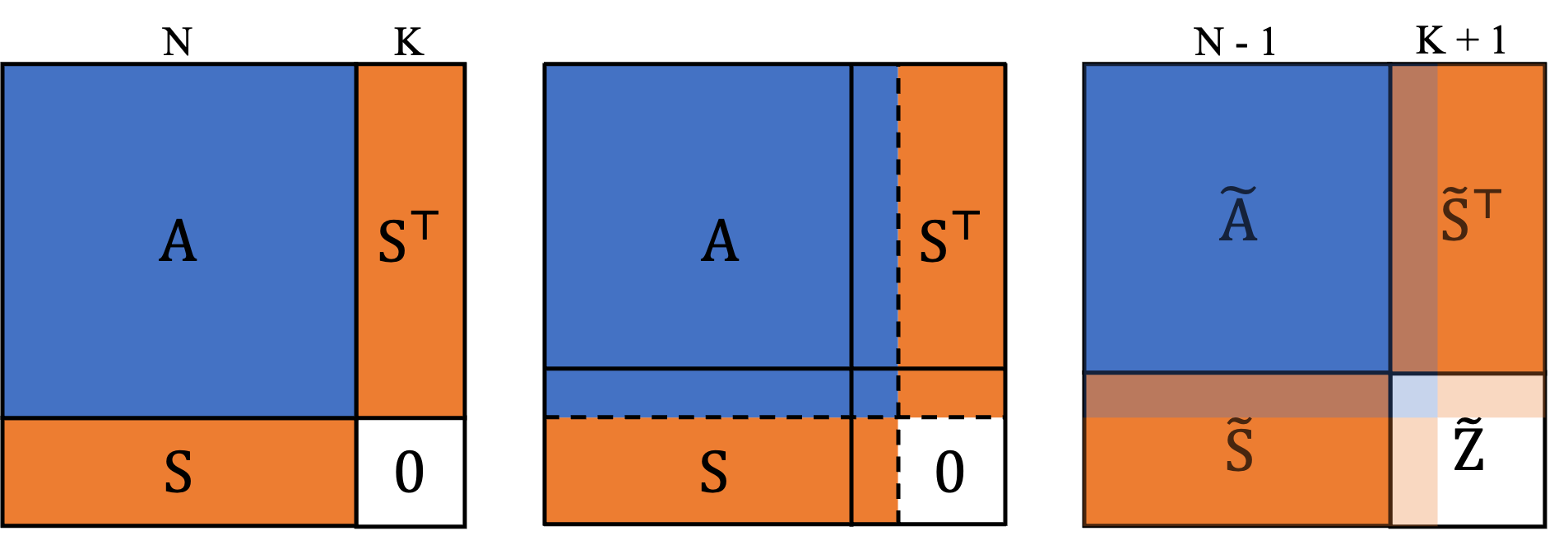}
\caption{
Shaving-off technique handling the singularity of the Bilaplacian matrix $\bA$. When assuming the last vertex is fixed as one of the control points, we define a new matrix $\tilde{\bA}$ by taking the last row and column from $\tilde{\bA}$ and also expand the selector matrix $\mathbf{S}$ and the zero matrix region. Since the Bilaplacian matrix is rank $N-1$, the new matrix $\tilde{\bA}$ has full rank, and thus the Schur complement trick in Sec.~\ref{sec:our_reformulation} can be used.
}
\label{fig:schur_trick}
\end{figure}

The reformulation of $\bM$ (Eq.~\ref{eq:final_m}) cannot be directly used in our case since the Bilaplacian matrix $\bA$ is a singular matrix.
One possible approach to handling the singularity of $\bA$ is to leverage the shaving-off technique by Jacobson~\cite{schur_trick} while fixing a single control point during the control point search.
Namely, when rewriting the optimization problem in Eq.~\ref{eq:biharmonics_optimization_objective} into a system of linear equations as follows:
\begin{equation}
    \begin{bmatrix}
        \bA & \bS^\top\\
        \bS & \mathbf{0}
    \end{bmatrix}
    \begin{bmatrix}
    \bX\\
    \Lambda
    \end{bmatrix}
    = 
    \begin{bmatrix}
    \mathbf{0}\\
    \bC
    \end{bmatrix},
\end{equation}
where $\Lambda \in \mathbb{R}^{K}$ is a vector of Lagrange multipliers, and assuming the fixed control point occupies the last index of vertices without loss of generality, the matrix on the left side of the linear system can be re-split in a way to shave off the last row and column of the Bilaplacian matrix $\mathbf{A}$ while expanding the selector matrix $\mathbf{S}$ and the zero matrix region, resulting in $\tilde{\bA}\in\mathbb{R}^{(N-1) \times (N-1)}$, $\tilde{\bS}\in\mathbb{R}^{(K+1) \times (N-1)}$, and $\tilde{\bZ}\in\mathbb{R}^{(K+1) \times (K+1)}$ as follows (see Fig.~\ref{fig:schur_trick}):
\begin{equation}
    \begin{bmatrix}
        \tilde{\bA} & \tilde{\bS}^\top\\
        \tilde{\bS} & \tilde{\bZ}
    \end{bmatrix}
    \begin{bmatrix}
    \bX\\
    \Lambda
    \end{bmatrix}
    = 
    \begin{bmatrix}
    \mathbf{0}\\
    \tilde{\bC}
    \end{bmatrix}.
\end{equation}

\noindent where $\tilde{\bC} \in \mathbb{R}^{(K+1) \times 3} $ is the concatenation of a zero vector and $\bC \in \mathbb{R}^{K \times 3}$. Since the Bilaplacian matrix $\bA$ has rank $N - 1$, $\tilde{\bA}$ obtained by removing the last row and column of $\bA$ has full rank. Thus, the Schur complement trick in Sec.~\ref{sec:our_reformulation} can now be directly utilized. One difference is that the bottom-right block matrix on the left side of the new linear system is not a zero matrix but $\tilde{\bZ}$. Hence, Eq.~\ref{eq:final_m} needs to be modified as follows:
\begin{equation}
    \bM = -\tilde{\bT} \tilde{\bA}^{-1} \tilde{\bS}^{\top}\left(\tilde{\bS} \tilde{\bA}^{-1} \tilde{\bS}^{\top} - \tilde{\bZ}\right)^{-1},
    \label{eq:final_m_new}
\end{equation}
where $\tilde{\bT}\in\{0, 1\}^{(N-K) \times (N-1)}$ is the complement of the selector matrix $\bT$ without the last column. Note that the last vertex represents the fixed control point, and thus, the complementary set of the control point selection in $\tilde{\bT}$ is not changed.
$\bW (\bS; \bA)$ in Eq.~\ref{eq:our_reformulation} is also reformulated again as follows:
\begin{equation}
    \tilde{\bW} (\bS; \bA) = \tilde{\bS}^{\top} + \tilde{\bT}^{\top}\left(\tilde{\bT} \tilde{\bA}^{-1} \tilde{\bS}^{\top}\left(\tilde{\bS} \tilde{\bA}^{-1} \tilde{\bS}^{\top} - \tilde{\bZ}\right)^{-1}\right).
    \label{eq:our_reformulation_new}
\end{equation}

\noindent The positions of the vertices, excluding the fixed single control point, can be computed as $\tilde{\mathbf{W}} \tilde{\mathbf{C}}$; note that the position of the control point is given.

Alternatively, one can simply consider regularizing the Bilaplacian matrix $\mathbf{A}$ by adding a small-weighted identity matrix (e.g., $\mathbf{A} + \epsilon \mathbf{I}$), approximating the solution while achieving numerical stability. 
In our experiments, we empirically find that this simple approach, which does not even require fixing any control points, performs well in practice for identifying the best set of control points. As a result, we use this regularization approach in our implementation.

\subsection{Control Point Search Algorithm}
\label{sec:search}

\begin{algorithm}[!t]
\caption{Pseudocode of ~\OptCtrlPoints{}.}
\label{alg:search_method} 
{
    \SetKw{Continue}{continue}
    \SetKwRepeat{Do}{do}{while}
    \footnotesize
    \tcc{Input: The ordered list of initial vertex indices, the ordered list of sets of the partitioned vertex indices, and the set of target shapes.}
    \tcc{Output: The ordered list of control point vertex indices.}
    \KwInput{$(\bs_k^{(0)})_{k=1}^{N}, \,\, ( V_k )_{k=1}^{N}, \,\, \{ \mathcal{X}_i \}_{i=1}^{M}.$}
    \KwOutput{$(\bs_k)_{k=1}^{N}$}

    \SetKwFunction{FD}{FittingDist}
    \SetKwFunction{FR}{FindRegion}
    \SetKwFunction{FV}{FindVertex}
    
    \SetKwProg{Fn}{Function}{:}{}
    \Fn{\FD{$( \bs_k )_{k=1}^{N}$}}{
        {$\bS \leftarrow \bS(( \bs_k )_{k=1}^{N})$\tcp*{Update the binary matrix.}}
        {$d \leftarrow \sum_i d(\mathcal{X}_i, \bW(\bS; \bA) \bC(\bS; \mathcal{X}_i))$\tcp*{Eq.~\ref{eq:loss}}}
        \KwRet $d$\;
    }
    
    \SetKwProg{Fn}{Function}{:}{}
    \Fn{\FR{$( \bs_k )_{k=1}^{N}, k, d_{\text{min}}$}}{
        $l_{\text{min}} \leftarrow k$\;

        \For{$l = 1, \ldots, K$} {
            \tcc{Resample $\bs^\prime$ if it is already selected.}
            \Do{$\bs^\prime \in \{ \bs_k \}_{k=1}^{N}$} {$\bs^\prime \sim U(V_l)$\tcp*{Draw a sample vertex.}}
            $\bs_k \leftarrow \bs^\prime$\;
            $d^\prime \leftarrow$ \FD($( \bs_k )_{k=1}^{N}$))\;
            \If { $d^\prime < d_{\text{min}}$ } {
                $d_{\text{min}} \leftarrow d^\prime$\;
                $l_{\text{min}} \leftarrow l$\;
            }
        }
        \KwRet $l_{\text{min}}, d_{\text{min}}$\;
    }

    \SetKwProg{Fn}{Function}{:}{}
    \Fn{\FV{$( \bs_k )_{k=1}^{N}, k, d_{\text{min}}, V$}}{
        $\bs^\prime_{\text{min}} \leftarrow s_k$\;

        \For{$\bs^\prime \in V$} {
            $\bs_k \leftarrow \bs^\prime$\;
            $d^\prime \leftarrow$ \FD($( \bs_k )_{k=1}^{N}$))\;
            \If { $d^\prime < d_{\text{min}}$ } {
                $d_{\text{min}} \leftarrow d^\prime$\;
                $\bs^\prime_{\text{min}} \leftarrow \bs^\prime$\;
            }
        }
        \KwRet $\bs^\prime_{\text{min}}, d_{\text{min}}$\;
    }

    $d_{\text{min}} \leftarrow \inf$

    \For{$k = 1, \ldots, K$} {
        $\bs_k \leftarrow \bs_k^{(0)}$\tcp*{Initialize with geodesic FPS.}
    }
    
    \For{$k = 1, \dots, K$} {
        $l, d_{\text{min}} \leftarrow$ \FR($( \bs_k )_{k=1}^{N}, k, d_{\text{min}}$))\;
        $\bs^\prime, d_{\text{min}} \leftarrow$ \FV($( \bs_k )_{k=1}^{N}, k, d_{\text{min}}, V_l$))\;
        $\bs_k \leftarrow \bs^\prime$\tcp*{Update the $k$-th control point.}
    }
}
\end{algorithm}

Although the computation of the biharmonic weight matrix $\mathbf{W}$ in Eq.~\ref{eq:our_reformulation} is fast, finding the $K$ optimal control points that best align the template mesh to the target shapes via deformation remains computationally infeasible when an exhaustive search of ${N \choose K}$ computations is used. To address this issue, we propose an effective search algorithm that reduces the time complexity to asymptotically linear order of the number of vertices $\Theta(N + K^2)$ in average. Despite this reduction in complexity, our algorithm still manages to discover nearly optimal solutions in practice.

In our search algorithm, our objective is to iteratively refine a set of control points starting from an initial configuration. We utilize geodesic Farthest Point Sampling (FPS) over surface of the mesh to establish the initial set. Our algorithm incorporates two key ideas:
\begin{itemize}
    \item Drawing inspiration from the coordinate descent approach in continuous optimization, we propose to determine the optimal location for each control point individually, while keeping all other current control points fixed.
    \item We propose a level-of-detail approach where at each iteration of updating a single control point, we select one of the partitioned volumes of the template mesh first and then traverse each vertex in the selected partition.
\end{itemize}
Specifically, let $( \mathbf{s}_k )_{k=1}^{N}$ denote the ordered list of template vertex indices for the control points, where $\bs_k \in [1, N]$ for all $k$, and $\bs_k \neq \mathbf{s}_l$ for all distinct $k$ and $l$. Let $\bS(( \bs_k )_{k=1}^{N})$ then represent the $K \times N$ binary matrix, with elements equal to one for the selected points and zero otherwise. The indices of the control points are initialized with the FPS point indices $( \bs_k^{(0)})_{k=1}^{N}$. We construct the partition of the vertex indices $\{V_k\}_{k=1}^{N}$ based on their proximity to the initial set of control points $( \bs_k^{(0)} )_{k=1}^{N}$, as shown inside the black boxes of Fig.~\ref{fig:pipeline}.
Since our algorithm allows the selection of internal vertices as control points, we employ the distance over the volume mesh graph as a measure of proximity.
We update each control point $\bs_k$ sequentially using the following two steps for each point. (See Alg.~\ref{alg:search_method} for the details.)

In the first step, we determine the partition to which the $k$-th control point $\bs_k$ will move. 
We randomly sample a vertex from each partitioned volume $V_l$. Then, we select the one of the sampled vertices that provides the minimum sum of distances between the template mesh and all the target shapes after deformation (as shown in Eq.~\ref{eq:loss}) when substituting $\bs_k$. By finding the vertex with the minimum sum of fitting distances, we identify the corresponding partitioned region $V$ for further exploration. (See \texttt{FindRegion} function in Alg.~\ref{alg:search_method}.) This approach allows each control point to explore different regions across the entire shape, mitigating the risk of falling into a local minimum through local search.

In the second step, within the selected region $V$, we find the best vertex, excluding those already chosen as control points, as a replacement for the $k$-th control point $\bs_k$ using the same distance measurement in Eq.~\ref{eq:loss}. The vertex selected during this step becomes the new $k$-th control point for the subsequent iteration. (See the \texttt{FindVertex} function in Alg. \ref{alg:search_method}.)

Assuming an even partitioning of the template mesh with an equal number of vertices in each region, the average time complexity to update the entire set of control points once is asymptotically $\mathcal{O}(K^2)$ for the first step and  $\Theta(N)$ for the second step. Therefore, the total complexity is $\Theta(N + K^2)$. This complexity is linear with respect to the number of vertices, and since $K \ll N$, it significantly reduces the computation time compared to the exhaustive search complexity of $\mathcal{O}(N^K)$.

\begin{figure*}[!ht]
\captionsetup{type=figure, labelfont=bf, textfont=it}
\begin{center}
\includegraphics[width=\textwidth]{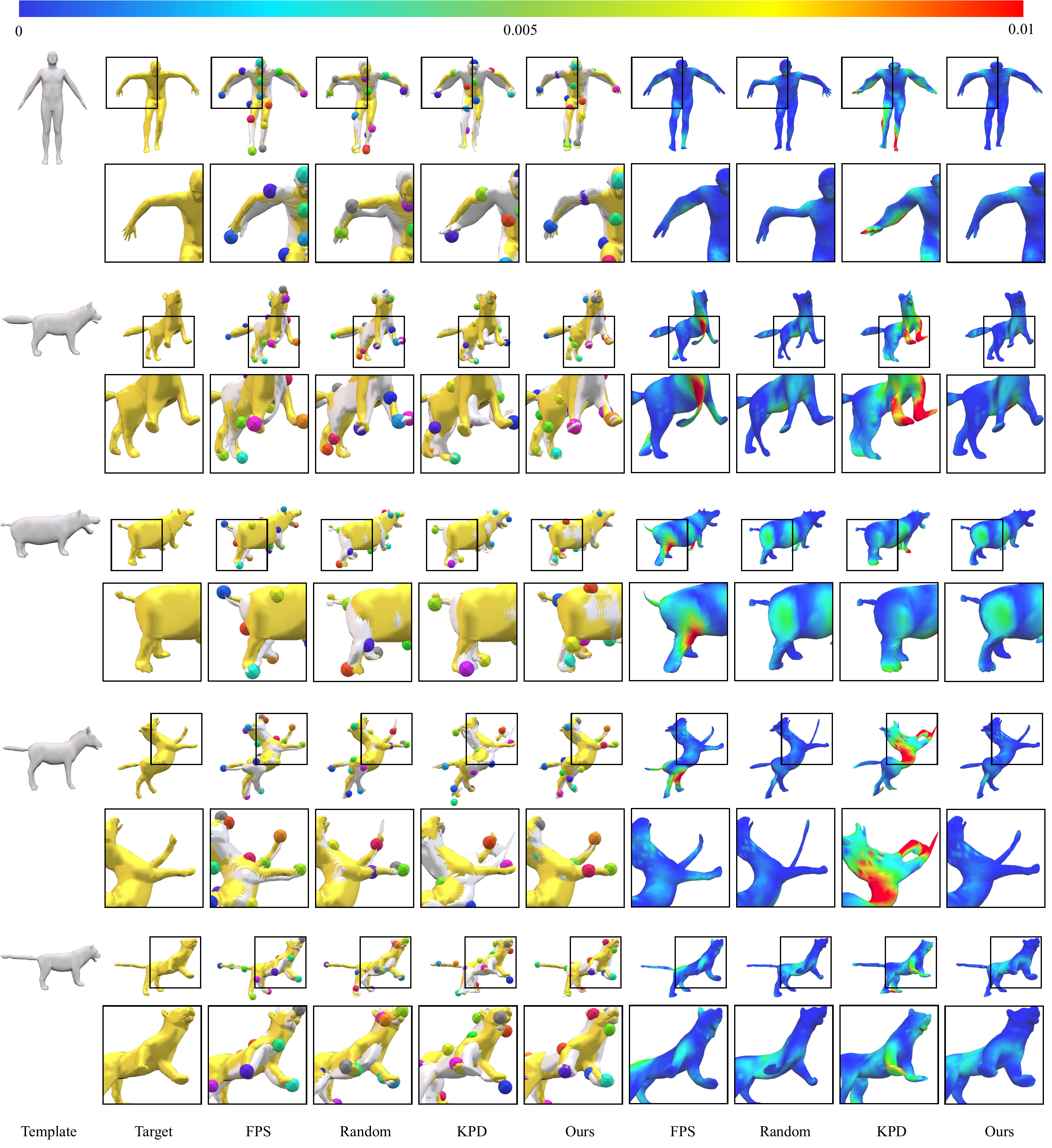}
\end{center}
\caption{
\textbf{Qualitative Results on the SMPL~\cite{SMPL:2015} and SMAL~\cite{Zuffi:CVPR:2017} datasets.} We show qualitative comparisons of our approach compared to FPS, random search and KPD, respectively. 
For each example: (Left) We show each method's output control points, the corresponding deformed template (white) overlayed over the desired target (yellow) to illustrate the alignment of the deformed source to the target shapes using the output control points.
Notice that our approach finds better control points near joints as shown in the shoulder of the human, legs of the fox, horse, hippo and tiger. 
(Right) We also show the raw output deformed source shape colored with vertex-to-vertex alignment to target error map for visualization. We see that apart from achieving better fitting error, our approach achieves less distortions especially on the limbs. 
In all examples, the bottom row is a zoomed in version of the top row. \textbf{Best viewed in zoom and color. Refer to the appendix for additional results.}
}
\label{fig:target-driven1}
\end{figure*}

\begin{figure*}[!ht]
\captionsetup{type=figure, labelfont=bf, textfont=it}
\begin{center}
\includegraphics[width=\textwidth]{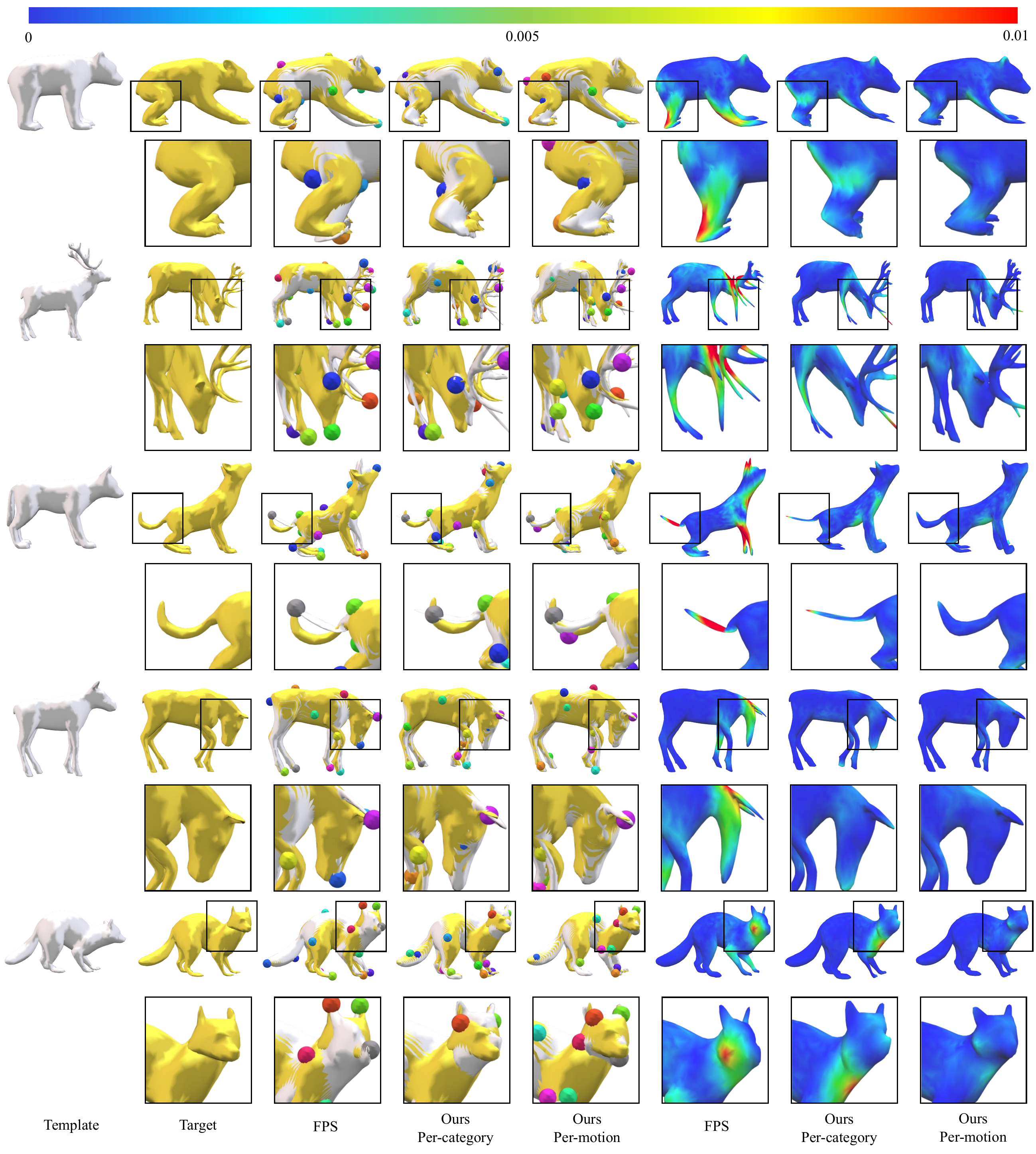}
\end{center}
\vspace{-0.2cm}
\caption{
\textbf{Qualitative Results on Deform4DThings~\cite{deformthings4d}.} We show qualitative comparisons of our approach in both per-category and per-motion settings compared to FPS. 
(Left) We show the output control points together with the corresponding deformed source shape (white) overlayed with the desired target (yellow). 
We see that our approach leads to better fitting compared to the FPS baseline. 
Moreover, our per-motion setting outputs more specialized control points that leads to better fitting to the specific motion, as shown by the legs of the bear (first row), head of the deer and raccoon (second and last row), tail of the doggie (third row), back and head of the horse (fourth row). 
(Right) We similarly also show a colored visualization of target shape that corresponds to the vertex-to-vertex error map.
We see significantly less distortions of our output shapes compared to the baseline.
\textbf{Best viewed in zoom and color. Refer to the appendix for additional results.}
}
\label{fig:target-driven2}
\end{figure*}

\section{Experiments}
\label{sec:results}
In this section, we present the results of our experiments, where we compare the performance of our proposed method, \OptCtrlPoints{}, with baseline search methods and a neural-network-based keypoint prediction method. We evaluate the performance based on the fitting error to the target shapes after deformation and the computation time.

\subsection{Datasets}
We evaluate our method on three different datasets of human (SMPL~\cite{SMPL:2015}) and animal (SMAL~\cite{Zuffi:CVPR:2017} and DeformingThings4D~\cite{deformthings4d}) models.
For each class of shapes, we take one template mesh and multiple target shapes covering a wide range of non-rigid deformations. 

\subsubsection{SMPL~\cite{SMPL:2015} and SMAL~\cite{Zuffi:CVPR:2017}} For human models,
we use synthetic shapes generated from SMPL~\cite{SMPL:2015}. We use the samples generated by Groueix \etal~\cite{3dcoded}, which contains a large variety of body poses and shapes.
For animal models,
we use four classes of shapes including \emph{fox}, \emph{hippo}, \emph{horse}, and \emph{tiger} generated from SMAL~\cite{Zuffi:CVPR:2017}.
We follow Groueix~\etal~\cite{3dcoded} to randomly draw the shape samples.
We run our searching algorithm on each animal category separately. For both SMPL and SMAL, we use the rest pose as the template shape and take $M=1000$ random shapes for each template as targets $\{\mathcal{X}_i\}_{i=1}^M$.

\subsubsection{DeformingThings4D~\cite{deformthings4d}} DeformingThings4D~\cite{deformthings4d} contains characters from Adobe Mixamo and also multiple animated motions for each character. 
We evaluate our approach both with \emph{all} motions for \emph{each} character and also with each motion to demonstrate the effectiveness of our \emph{data-driven} method of locating the ideal set of control points best fitting the given set of targets.
We use seven characters in our experiments, namely \emph{bear}, \emph{deer}, \emph{doggie}, \emph{dragon}, \emph{moose}, \emph{procy} and \emph{raccoon}.
For each character, we randomly sample $M=1000$ different targets from all motions for the per-category experiments, while we use all the frames of the motion as targets for the per-motion experiments. We use the first frame of specific animation sequence as the template shape. Refer to the appendix for more details about the data used in our experiments.

\subsection{Experiment Setup}

\subsubsection{Data Preprocessing and Implementation Details} For all template meshes, we first convert each mesh into a watertight manifold using the method of Huang~\etal~\cite{Huang2018ARXIV} and then into a tetrahedral mesh using TetWild~\cite{Hu2018SIGGRAPH}. We simplify and regularize each tetrahedral mesh to have 5000 vertices and also normalize it to fit in a unit sphere. We then precompute the Bilaplacian matrix $\mathbf{A}$ for each template mesh using the libigl~\cite{libigl} and and its inverse with the regularization described in Sec.~\ref{sec:singularity} for efficient searching.

To compute $\bC(\bS; \mathcal{X}_i) = \bS f(\mathcal{X}_i)$ (Sec.~\ref{sec:problem_definition}), we leverage vertex-wise correspondence of the meshes provided from SMPL~\cite{SMPL:2015}, SMAL~\cite{Zuffi:CVPR:2017}, and DeformingThings4D~\cite{deformthings4d}, while the correspondence can also be found using the off-the-shelf shape correspondence methods (refer to a survey~\cite{shape_corr_survey} for the recent literature).
Given the vertex-wise correspondence, we use the average of per-vertex L2 distance as our shape-to-shape distance function $d(\cdot, \cdot)$ (Sec.~\ref{sec:problem_definition}). We executed Alg.~\ref{alg:search_method} only once in all the experiments, but we also demonstrate in Sec.~\ref{sec:refine_iteration} that iterating the algorithm further improves the selection of control points.

\subsubsection{Baselines}
We compare our method, \OptCtrlPoints{}, with three different baselines:

\begin{enumerate}
\item \textbf{Farthest Point Sampling (FPS)}: This is the case of directly using the geodesic Farthest Point Sampling vertices, our initial set of control points $( \bs_k^{(0)})_{k=1}^{N}$, as the final set without searching.
We demonstrate in our experiments that our efficient searching method discovers a much better set of control points, which greatly reduces the fitting error.
\item \textbf{Random Search}: Instead of the computationally infeasible exhaustive search that tests all possible ${N \choose K}$ cases, we compare our method with a random search approach. In random search, we randomly select $K$ control points from the $N$ vertices multiple times and choose the set with the lowest fitting distance. We iterate the random set sampling process $N \times K$ times, which is significantly larger than the number of cases in our method.
\item \textbf{KeypointDeformer (KPD)~\cite{jakab2021keypointdeformer}}: We additionally compare our method with KPD, which leverages a neural-network-based approach to predict keypoints on shapes and align the template to the target through cage-based deformation (not biharmonic). We trained KPD for each template shape and its corresponding set of $M$ targets using the released codebase. The cage created by warping a \emph{sphere} mesh often fails to disentangle the deformations of different parts, especially in shapes with articulating parts (such as limbs of a human body) or complex topological structures. In our experiments, we demonstrate that our method, employing biharmonic deformation, achieves a better fit of the template shape to the targets with the ideal set of control points.
\end{enumerate}

\begin{table}[t!]
\captionsetup{type=table, labelfont=bf, textfont=it}
{\footnotesize
\begin{tabularx}{\columnwidth}{YYYYYYY}
\toprule
\multicolumn{7}{c}{Average Fitting Distance ($\times 10^{-4}$)} \\ \midrule
\multicolumn{1}{c|}{\multirow{2}{*}{Methods}} &
  \multicolumn{1}{c|}{\multirow{2}{*}{K}} &
  \multicolumn{1}{c|}{\multirow{2}{*}{\makecell{SMPL\\\cite{SMPL:2015}}}} &
  \multicolumn{4}{c}{SMAL\cite{Zuffi:CVPR:2017}} \\ \cmidrule{4-7} 
\multicolumn{1}{c|}{} &
  \multicolumn{1}{c|}{} &
  \multicolumn{1}{c|}{} &
  \multicolumn{1}{c|}{fox} &
  \multicolumn{1}{c|}{hippo} &
  \multicolumn{1}{c|}{horse} &
  tiger \\ \midrule
\multicolumn{1}{c|}{\multirow{3}{*}{\makecell{KPD\\\cite{jakab2021keypointdeformer}}}} &
  \multicolumn{1}{c|}{16} &
  \multicolumn{1}{c|}{17.09} &
  \multicolumn{1}{c|}{23.12} &
  \multicolumn{1}{c|}{23.41} &
  \multicolumn{1}{c|}{34.82} &
  21.53 \\
\multicolumn{1}{c|}{} &
  \multicolumn{1}{c|}{24} &
  \multicolumn{1}{c|}{17.16} &
  \multicolumn{1}{c|}{25.63} &
  \multicolumn{1}{c|}{23.65} &
  \multicolumn{1}{c|}{28.42} &
  22.92 \\
\multicolumn{1}{c|}{} &
  \multicolumn{1}{c|}{32} &
  \multicolumn{1}{c|}{14.49} &
  \multicolumn{1}{c|}{25.70} &
  \multicolumn{1}{c|}{20.75} &
  \multicolumn{1}{c|}{32.47} &
  20.15 \\ \midrule
\multicolumn{1}{c|}{\multirow{3}{*}{FPS}} &
  \multicolumn{1}{c|}{16} &
  \multicolumn{1}{c|}{11.28} &
  \multicolumn{1}{c|}{10.16} &
  \multicolumn{1}{c|}{12.57} &
  \multicolumn{1}{c|}{10.37} &
  8.47 \\
\multicolumn{1}{c|}{} &
  \multicolumn{1}{c|}{24} &
  \multicolumn{1}{c|}{5.70} &
  \multicolumn{1}{c|}{4.41} &
  \multicolumn{1}{c|}{8.32} &
  \multicolumn{1}{c|}{3.92} &
  3.76 \\
\multicolumn{1}{c|}{} &
  \multicolumn{1}{c|}{32} &
  \multicolumn{1}{c|}{3.60} &
  \multicolumn{1}{c|}{2.97} &
  \multicolumn{1}{c|}{5.38} &
  \multicolumn{1}{c|}{2.52} &
  2.60 \\ \midrule
\multicolumn{1}{c|}{\multirow{3}{*}{\makecell{Random\\Search}}} &
  \multicolumn{1}{c|}{16} &
  \multicolumn{1}{c|}{8.38} &
  \multicolumn{1}{c|}{8.18} &
  \multicolumn{1}{c|}{9.20} &
  \multicolumn{1}{c|}{7.88} &
  6.78 \\
\multicolumn{1}{c|}{} &
  \multicolumn{1}{c|}{24} &
  \multicolumn{1}{c|}{3.91} &
  \multicolumn{1}{c|}{4.78} &
  \multicolumn{1}{c|}{5.38} &
  \multicolumn{1}{c|}{4.52} &
  3.80 \\
\multicolumn{1}{c|}{} &
  \multicolumn{1}{c|}{32} &
  \multicolumn{1}{c|}{2.51} &
  \multicolumn{1}{c|}{3.40} &
  \multicolumn{1}{c|}{3.63} &
  \multicolumn{1}{c|}{2.97} &
  2.59 \\ \midrule
\multicolumn{1}{c|}{\multirow{3}{*}{Ours}} &
  \multicolumn{1}{c|}{16} &
  \multicolumn{1}{c|}{\textbf{5.16}} &
  \multicolumn{1}{c|}{\textbf{5.07}} &
  \multicolumn{1}{c|}{\textbf{5.49}} &
  \multicolumn{1}{c|}{\textbf{4.45}} &
  \textbf{4.15} \\
\multicolumn{1}{c|}{} &
  \multicolumn{1}{c|}{24} &
  \multicolumn{1}{c|}{\textbf{2.25}} &
  \multicolumn{1}{c|}{\textbf{2.61}} &
  \multicolumn{1}{c|}{\textbf{3.06}} &
  \multicolumn{1}{c|}{\textbf{2.13}} &
  \textbf{2.21} \\
\multicolumn{1}{c|}{} &
  \multicolumn{1}{c|}{32} &
  \multicolumn{1}{c|}{\textbf{1.36}} &
  \multicolumn{1}{c|}{\textbf{1.89}} &
  \multicolumn{1}{c|}{\textbf{1.83}} &
  \multicolumn{1}{c|}{\textbf{1.57}} &
  \textbf{1.42} \\ \bottomrule
\end{tabularx}%
}
\centering
\caption{Average fitting distance between corresponding vertices of target shape and deformed template shape multiplied by $10^{4}$.}
\label{tab:target-driven1}
\end{table}

\subsection{Fitting Error Comparisons with SMPL~\cite{SMPL:2015} and SMAL~\cite{Zuffi:CVPR:2017}}
\label{sec:SMPL_SMAL_results}

We evaluate our~\OptCtrlPoints{} and other baselines by assessing how well the template mesh aligns with the target shapes using the output deformation handles.
We conducted experiments with all the methods while varying the number of keypoints $K$ to be 16, 24, and 32.

First, we compare our method to FPS and random search, where control points are used as handles for biharmonic deformation. Tab.~\ref{tab:target-driven1} presents a comparison of the average fitting distances between the template and target shapes. Compared to FPS, where our initial set of control points is used directly, our method demonstrates a significant improvement, reducing the fitting distance by more than half in most cases. Additionally, when compared to random search, which explores a much larger number of control point sets, our efficient search yields substantially lower fitting errors.

Qualitative results in Fig.~\ref{fig:target-driven1} also show that our method achieves more meaningful fine-grained deformations than FPS and random search, thanks to the optimal placement of control points in regions with greater variations or articulations. 
The first column shows the templates, the second column displays the targets, the next four columns demonstrate the alignment results through deformation, and the last four columns exhibit the fitting error maps over the deformed template shapes. 
Notably, failure cases of FPS and random search are observed, resulting in distortion in the deformation, such as in the \emph{second row} where the legs of the fox are distorted, and in the \emph{last row} where the legs and body of the tiger deviates from the target shapes. In contrast, our method's control points produce much better deformation results.

Furthermore, compared to KPD~\cite{jakab2021keypointdeformer}, which employs cage-based deformation instead of biharmonic deformation, our method excels in fitting the template to the targets through deformation. 
Tab.~\ref{tab:target-driven1} clearly illustrates a substantial gap between the average fitting errors of KPD and our method. 
Moreover, Fig.~\ref{fig:target-driven1} vividly showcases the qualitative difference, especially in the arms and legs of the human (\emph{first row}) and the hind legs of the fox and the horse (\emph{second} and \emph{fourth rows}).

\begin{table}[t!]
\captionsetup{type=table, labelfont=bf, textfont=it}
{
\footnotesize
{%
\begin{tabularx}{0.8\columnwidth}{c|Y|Y|Y}
\toprule
\multirow{2}{*}{Methods}  &  \multicolumn{3}{c}{Time (mins)}     \\ \cmidrule{2-4}
& $K=16$ & $K=24$ & $K=32$\\
\midrule
Random Search & 48.8 & 70.4 & 97.0 \\
\midrule
Ours & \textbf{2.8} & \textbf{2.9} & \textbf{3.0} \\
\bottomrule
\end{tabularx}%
}
\centering
\caption{Execution time profiling of our \OptCtrlPoints{} compared to random search. We show that our method demonstrates significantly faster performance compared to random search as our time complexity is reduced to $\Theta(N+K^2)$, whereas the time requirement increase linearly with the number of control points for random search.}
\label{tab:exp_time}
}
\end{table}

\textbf{Figures are best viewed with zoom and in color. Additional qualitative results can be found in the supplementary material}.

\begin{table*}[t!]
\captionsetup{type=table, labelfont=bf, textfont=it}
{\footnotesize
\resizebox{2.1\columnwidth}{!}{%
\begin{tabular}{@{}cccccccccccccccc@{}}
\toprule
\multicolumn{16}{c}{Average Fitting Distance ($\times 10^{-4}$)} \\ \midrule
\multicolumn{1}{c|}{\multirow{2}{*}{Methods}} &
  \multicolumn{1}{c|}{\multirow{2}{*}{K}} &
  \multicolumn{2}{c|}{Bear (3EP)} &
  \multicolumn{2}{c|}{Deer (OMG)} &
  \multicolumn{2}{c|}{Doggie (MN5)} &
  \multicolumn{2}{c|}{Dragon (OF2)} &
  \multicolumn{2}{c|}{Moose (1DOG)} &
  \multicolumn{2}{c|}{Procy (STEM)} &
  \multicolumn{2}{c}{Raccoon (VGG)} \\ \cmidrule(lr){3-16}
\multicolumn{1}{c|}{} &
  \multicolumn{1}{c|}{} &
  Cat. &
  \multicolumn{1}{c|}{Mot.} &
  Cat. &
  \multicolumn{1}{c|}{Mot.} &
  Cat. &
  \multicolumn{1}{c|}{Mot.} &
  Cat. &
  \multicolumn{1}{c|}{Mot.} &
  Cat. &
  \multicolumn{1}{c|}{Mot.} &
  Cat. &
  \multicolumn{1}{c|}{Mot.} &
  Cat. &
  Mot. \\ \midrule
\multicolumn{1}{c|}{\multirow{3}{*}{FPS}} &
  \multicolumn{1}{c|}{16} &
  18.79 &
  \multicolumn{1}{c|}{16.06} &
  15.89 &
  \multicolumn{1}{c|}{17.40} &
  20.13 &
  \multicolumn{1}{c|}{16.97} &
  45.33 &
  \multicolumn{1}{c|}{44.35} &
  14.72 &
  \multicolumn{1}{c|}{14.99} &
  26.24 &
  \multicolumn{1}{c|}{24.57} &
  30.83 &
  27.88 \\  
\multicolumn{1}{c|}{} &
  \multicolumn{1}{c|}{24} &
  8.78 &
  \multicolumn{1}{c|}{7.72} &
  5.33 &
  \multicolumn{1}{c|}{5.59} &
  7.48 &
  \multicolumn{1}{c|}{6.24} &
  26.35 &
  \multicolumn{1}{c|}{26.31} &
  6.66 &
  \multicolumn{1}{c|}{6.57} &
  15.24 &
  \multicolumn{1}{c|}{14.40} &
  15.68 &
  15.65 \\  
\multicolumn{1}{c|}{} &
  \multicolumn{1}{c|}{32} &
  6.82 &
  \multicolumn{1}{c|}{6.05} &
  3.10 &
  \multicolumn{1}{c|}{3.14} &
  6.32 &
  \multicolumn{1}{c|}{5.31} &
  16.85 &
  \multicolumn{1}{c|}{17.06} &
  4.94 &
  \multicolumn{1}{c|}{4.86} &
  9.89 &
  \multicolumn{1}{c|}{9.70} &
  10.87 &
  11.35 \\ \midrule
\multicolumn{1}{c|}{\multirow{3}{*}{\makecell{Random\\Search}}} &
  \multicolumn{1}{c|}{16} &
  16.17 &
  \multicolumn{1}{c|}{12.59} &
  8.35 &
  \multicolumn{1}{c|}{7.14} &
  14.76 &
  \multicolumn{1}{c|}{11.27} &
  21.57 &
  \multicolumn{1}{c|}{19.91} &
  14.04 &
  \multicolumn{1}{c|}{12.25} &
  20.16 &
  \multicolumn{1}{c|}{17.86} &
  27.75 &
  24.99 \\  
\multicolumn{1}{c|}{} &
  \multicolumn{1}{c|}{24} &
  9.83 &
  \multicolumn{1}{c|}{7.77} &
  3.74 &
  \multicolumn{1}{c|}{3.32} &
  8.63 &
  \multicolumn{1}{c|}{6.65} &
  16.04 &
  \multicolumn{1}{c|}{15.41} &
  6.77 &
  \multicolumn{1}{c|}{6.37} &
  12.19 &
  \multicolumn{1}{c|}{10.36} &
  16.54 &
  15.12 \\  
\multicolumn{1}{c|}{} &
  \multicolumn{1}{c|}{32} &
  5.95 &
  \multicolumn{1}{c|}{5.05} &
  2.54 &
  \multicolumn{1}{c|}{2.31} &
  5.86 &
  \multicolumn{1}{c|}{4.56} &
  14.02 &
  \multicolumn{1}{c|}{13.50} &
  4.52 &
  \multicolumn{1}{c|}{4.17} &
  7.82 &
  \multicolumn{1}{c|}{6.92} &
  11.96 &
  11.23 \\ \midrule
\multicolumn{1}{c|}{\multirow{3}{*}{Ours}} &
  \multicolumn{1}{c|}{16} &
  9.67 &
  \multicolumn{1}{c|}{\textbf{7.65}} &
  5.23 &
  \multicolumn{1}{c|}{\textbf{4.19}} &
  8.72 &
  \multicolumn{1}{c|}{\textbf{6.67}} &
  17.80 &
  \multicolumn{1}{c|}{\textbf{16.76}} &
  6.62 &
  \multicolumn{1}{c|}{\textbf{6.37}} &
  13.14 &
  \multicolumn{1}{c|}{\textbf{11.10}} &
  16.13 &
  \textbf{14.56} \\  
\multicolumn{1}{c|}{} &
  \multicolumn{1}{c|}{24} &
  5.16 &
  \multicolumn{1}{c|}{\textbf{4.06}} &
  1.72 &
  \multicolumn{1}{c|}{\textbf{1.56}} &
  4.57 &
  \multicolumn{1}{c|}{\textbf{3.54}} &
  10.68 &
  \multicolumn{1}{c|}{\textbf{10.31}} &
  3.78 &
  \multicolumn{1}{c|}{\textbf{3.51}} &
  6.46 &
  \multicolumn{1}{c|}{\textbf{5.50}} &
  9.10 &
  \textbf{8.42} \\  
\multicolumn{1}{c|}{} &
  \multicolumn{1}{c|}{32} &
  3.47 &
  \multicolumn{1}{c|}{\textbf{2.73}} &
  1.28 &
  \multicolumn{1}{c|}{\textbf{1.15}} &
  2.95 &
  \multicolumn{1}{c|}{\textbf{2.31}} &
  9.29 &
  \multicolumn{1}{c|}{\textbf{8.95}} &
  2.74 &
  \multicolumn{1}{c|}{\textbf{2.44}} &
  4.35 &
  \multicolumn{1}{c|}{\textbf{3.73}} &
  5.80 &
  \textbf{5.37} \\ \bottomrule
\end{tabular}%
}
\centering
\caption{Target-driven shape deformation results for DeformingThings4D~\cite{deformthings4d} dataset.
The control points identified by our \OptCtrlPoints{} method achieve better alignment of the template to the targets compared to FPS and random search. Moreover, when specific motion targets (per-motion, denoted as Mot.) are provided instead of general targets across all motions (per-category, denoted as Cat.), the control points are further tailored, resulting in even lower fitting distances.
The L2-loss across corresponding vertices is multiplied by $10^{4}$.}
\label{tab:target-driven2}
\vspace{-1.5\baselineskip}
}
\end{table*}

\begin{table}[t!]
\captionsetup{type=table, labelfont=bf, textfont=it}
{
\footnotesize
{%
\begin{tabularx}{0.8\columnwidth}{c|Y|Y|Y}
\toprule
\multirow{2}{*}{Iteration}  &  \multicolumn{3}{c}{SMPL\cite{SMPL:2015}}     \\ \cmidrule{2-4}
& $K=16$ & $K=24$ & $K=32$\\
\midrule
1 & 5.14 & 2.31	& 1.34 \\
\midrule
2 & 4.77 & 1.95	& 1.17 \\
\midrule
3 & 4.74 & 1.80 & 1.15 \\
\bottomrule
\end{tabularx}%
}
\centering
\caption{Results from iterating Alg. \ref{alg:search_method}. Fitting distance is improved through iterative execution of Alg. \ref{alg:search_method} on the SMPL dataset across different numbers of control points. The L2-loss across corresponding vertices is multiplied by $10^{4}$.}
\label{tab:iteration}
\vspace{-1.5\baselineskip}
}
\end{table}

\subsection{Computation Time Analysis}
Our proposed reformulation for the biharmonic weights matrix computation (Sec.~\ref{sec:our_reformulation}) and the efficient search algorithm (Sec.~\ref{sec:search}) enable us to determine the optimal control point locations in a matter of minutes, even when dealing with 1000 target shapes. Without both of these advancements, achieving this task computationally would be challenging.

We first present profiling results comparing the computation time of the fitting loss (Eq.~\ref{eq:loss}) using the original biharmonic weight formation (Eq.~\ref{eq:biharmonics_solution}) and our reformulation (Eq.~\ref{eq:our_reformulation}). When computing the fitting loss with 16 control points, utilizing a single NVIDIA RTX 3090 and parallelization, our PyTorch implementation takes 1.188 seconds for the original formulation, while our reformulation with precomputation completes the calculation in only 0.024 seconds, which is \textbf{49 times faster} than the original formulation.

We also demonstrate the efficiency of our method compared to the naive random search approach for finding a better solution. 
We conducted profiling to compare the overall execution time of our \OptCtrlPoints{} and random search. 
Tab.~\ref{tab:exp_time} presents the average runtime of~\OptCtrlPoints{} compared to random search. In random search, we sample subsets of vertices $N\times K$ times, whereas our method has a linear order complexity with respect to the number of vertices. 
Consequently, we achieve approximately $K$ times more speedup, as shown in Tab.~\ref{tab:exp_time}, resulting in a significant reduction in computation time from about an hour to approximately \textbf{3 minutes}, while also obtaining a better set of control points (as demonstrated in the quantitative results in Tab.~\ref{tab:target-driven1}).

These findings emphasize the crucial role played by both the new biharmonic weight formation and the efficient search algorithm, as \textbf{without them, it would take days} to find a satisfactory set of control points. For instance, when the number of control points $K$ is 16, $48.8\,\text{mins} \times 49 = 2391.2\,\text{mins} = 1.66\,\text{days}$.

\subsection{Results with DeformingThings4D~\cite{deformthings4d}}
\label{sec:Deform4D_results}
In the experiment with the DeformThings4D dataset, we highlight the effectiveness of our~\OptCtrlPoints{} in discovering the optimal set of control points for a given set of targets in a \emph{data-driven} manner. 
We present two distinct experimental setups: per-category and per-motion.

Quantitative results are presented in Table~\ref{tab:target-driven2}, where our approach outperforms both FPS and random search in both the per-category and per-motion setups. 
Notably, our \OptCtrlPoints{} consistently achieves a lower fitting error in the per-motion setup compared to the per-category setup. 
This outcome is attributed to our \emph{data-driven} approach, which enables us to discover control points that are tailored to the given set of targets.

Fig.~\ref{fig:target-driven2} presents qualitative results. In the \emph{third row}, FPS fails to preserve the geometry of the dog's legs and tail, while our approach successfully retains the leg geometry during the howl motion by identifying control points on the joints.
Moreover, our per-motion approach significantly outperforms the FPS baseline in scenarios involving larger motions and wider variations in the targets. For example, in the \emph{second row}, our method achieves much better results in recovering the geometry of the deer's face and legs, whereas FPS falls short in this regard.
Also, in the \emph{fifth row}, our approach excels at capturing head deformations by identifying additional control points on the raccoon's head and neck.

\textbf{Figures are best viewed with zoom and in color. Additional qualitative results can be found in the supplementary material.}

\subsection{Refinement through Iteration of Algorithm \ref{alg:search_method}}
\label{sec:refine_iteration}

We show that while executing Algorithm \ref{alg:search_method} only once can achieve desirable results, iterating through the algorithm can yield improved outcomes, as shown in Tab. \ref{tab:iteration}. We see that through iterative execution of the Alg. \ref{alg:search_method}, the average fitting distance is further reduced, leading to better results on the SMPL dataset across different numbers of control points.

\section{Conclusion}
\label{sec:conclude}

We introduced \OptCtrlPoints{}, a data-driven method for determining the optimal set of control points to replicate target shapes as biharmonic deformations of the template mesh. To address the computational challenges associated with finding the best $k$-subset out of $N$ vertices while solving a large-scale linear system at each trial, we proposed a reformulation of the biharmonic weights that significantly speeds up the computation. Additionally, we developed an efficient search algorithm that significantly outperforms random search in terms of both quality and time efficiency.
In future work, we plan to extend our method to identify {\em region}  handles of 3D shapes. This extension will allow us to handle more localized and specific deformations in a controlled and data-driven manner.

\noindent \textbf{Acknowledgements.} This work was partly supported by NRF grant (RS-2023-00209723) and IITP grant (2022-0-00594, RS-2023-00227592) funded by the Korean government (MSIT), Technology Innovation Program (20016615) funded by the Korea
government(MOTIE), and grants from ETRI, KT, NCSOFT, and Samsung Electronics. Leonidas Guibas acknowledges support from an ARL grant W911NF-21-2-0104, a Vannevar Bush Faculty Fellowship, and gifts from the Adobe and Snap corporations. Despoina Paschalidou acknowledges support from the Swiss National Science Foundation under grant number P500PT 206946.

{\small
\bibliographystyle{eg-alpha-doi}
\bibliography{main}

\newcommand{\etalchar}[1]{$^{#1}$}
\begin{thebibliography}{\uppercase{MRSHO22}}

\bibitem[APO21]{attaiki2021dpfm}
\textsc{Attaiki S., Pai G., Ovsjanikov M.}:
\newblock {DPFM}: Deep partial functional maps.
\newblock In \emph{3DV} (2021).

\bibitem[BP07]{baran2007automatic}
\textsc{Baran I., Popovi\'{c} J.}:
\newblock Automatic rigging and animation of {3D} characters.
\newblock \emph{ACM Transactions on Graphics} (2007).

\bibitem[BPGK06]{botsch06}
\textsc{Botsch M., Pauly M., Gross M.~H., Kobbelt L.}:
\newblock {PriMo}: coupled prisms for intuitive surface modeling.
\newblock In \emph{Eurographics Symposium on Geometry Processing} (2006).

\bibitem[BPWG07]{bostch2007}
\textsc{Botsch M., Pauly M., Wicke M., Gross M.}:
\newblock Adaptive space deformations based on rigid cells.
\newblock \emph{Computer Graphics Forum} (2007).

\bibitem[BS08]{botsch08}
\textsc{Botsch M., Sorkine O.}:
\newblock On linear variational surface deformation methods.
\newblock \emph{IEEE Transactions on Visualization and Computer Graphics} (2008).

\bibitem[DSO20]{donati2020deep}
\textsc{Donati N., Sharma A., Ovsjanikov M.}:
\newblock {Deep Geometric Functional Maps}: Robust feature learning for shape correspondence.
\newblock In \emph{CVPR} (2020).

\bibitem[DYT21]{deng2021deformed}
\textsc{Deng Y., Yang J., Tong X.}:
\newblock {Deformed Implicit Field}: Modeling {3D} shapes with learned dense correspondence.
\newblock In \emph{CVPR} (2021).

\bibitem[GFK{\etalchar{*}}18]{3dcoded}
\textsc{Groueix T., Fisher M., Kim V.~G., Russell B., Aubry M.}:
\newblock {3D-CODED} : {3D} correspondences by deep deformation.
\newblock In \emph{ECCV} (2018).

\bibitem[GFK{\etalchar{*}}19]{CycleConsistency}
\textsc{Groueix T., Fisher M., Kim V.~G., Russell B.~C., Aubry M.}:
\newblock Deep self-supervised cycle-consistent deformation for few-shot shape segmentation.
\newblock In \emph{Eurographics Symposium on Geometry Processing} (2019).

\bibitem[Hay68]{haynsworth1968schur}
\textsc{Haynsworth E.~V.}:
\newblock \emph{On the Schur complement}.
\newblock Tech. rep., BASEL UNIV (SWITZERLAND) MATHEMATICS INST, 1968.

\bibitem[HFW{\etalchar{*}}18]{Hanocka:2018}
\textsc{Hanocka R., Fish N., Wang Z., Giryes R., Fleishman S., Cohen-Or D.}:
\newblock {ALIGNet}: Partial-shape agnostic alignment via unsupervised learning.
\newblock \emph{ACM Transactions on Graphics} (2018).

\bibitem[HLR{\etalchar{*}}19]{halimi2019unsupervised}
\textsc{Halimi O., Litany O., Rodola E., Bronstein A.~M., Kimmel R.}:
\newblock Unsupervised learning of dense shape correspondence.
\newblock In \emph{CVPR} (2019).

\bibitem[HS08]{Hormann2008MaximumEC}
\textsc{Hormann K., Sukumar N.}:
\newblock Maximum entropy coordinates for arbitrary polytopes.
\newblock \emph{Computer Graphics Forum} (2008).

\bibitem[HSG18]{Huang2018ARXIV}
\textsc{Huang J., Su H., Guibas L.}:
\newblock Robust watertight manifold surface generation method for shapenet models.
\newblock \emph{CoRR abs/1802.01698} (2018).

\bibitem[HZG{\etalchar{*}}18]{Hu2018SIGGRAPH}
\textsc{Hu Y., Zhou Q., Gao X., Jacobson A., Zorin D., Panozzo D.}:
\newblock Tetrahedral meshing in the wild.
\newblock \emph{SIGGRAPH} (2018).

\bibitem[IMH05]{Igarashi:2005}
\textsc{Igarashi T., Moscovich T., Hughes J.~F.}:
\newblock As-rigid-as-possible shape manipulation.
\newblock In \emph{SIGGRAPH} (2005).

\bibitem[Jac14]{schur_trick}
\textsc{Jacobson A.}:
\newblock \emph{Schur Complement Trick for Positive Semi-definite Energies}.
\newblock Tech. rep., Columbia University, 2014.

\bibitem[JBPS11]{biharmonic}
\textsc{Jacobson A., Baran I., Popovic J., Sorkine O.}:
\newblock Bounded biharmonic weights for real-time deformation.
\newblock In \emph{SIGGRAPH Asia} (2011).

\bibitem[JMD{\etalchar{*}}07]{Joshi:2007}
\textsc{Joshi P., Meyer M., DeRose T., Green B., Sanocki T.}:
\newblock Harmonic coordinates for character articulation.
\newblock In \emph{SIGGRAPH} (2007).

\bibitem[JP{\etalchar{*}}18]{libigl}
\textsc{Jacobson A., Panozzo D., et~al.}:
\newblock {libigl}: A simple {C++} geometry processing library, 2018.
\newblock https://libigl.github.io/.

\bibitem[JPS{\etalchar{*}}18]{Jack:2018}
\textsc{Jack D., Pontes J.~K., Sridharan S., Fookes C., Shirazi S., Maire F., Eriksson A.}:
\newblock Learning free-form deformations for {3D} object reconstruction.
\newblock In \emph{ICCV} (2018).

\bibitem[JSW05]{Ju_meanvalue}
\textsc{Ju T., Schaefer S., Warren J.}:
\newblock Mean value coordinates for closed triangular meshes.
\newblock \emph{ACM Transactions on Graphics} (2005).

\bibitem[JTM{\etalchar{*}}20]{jakab2021keypointdeformer}
\textsc{Jakab T., Tucker R., Makadia A., Wu J., Snavely N., Kanazawa A.}:
\newblock {KeypointDeformer}: Unsupervised {3D} keypoint discovery for shape control.
\newblock In \emph{CVPR} (2020).

\bibitem[KJG{\etalchar{*}}18]{Kurenkov:2018}
\textsc{Kurenkov A., Ji J., Garg A., Mehta V., Gwak J., Choy C.~B., Savarese S.}:
\newblock {DeformNet}: Free-form deformation network for {3D} shape reconstruction from a single image.
\newblock In \emph{WACV} (2018).

\bibitem[KSSCO06]{Kraevoy:2008}
\textsc{Kraevoy V., Sheffer A., Shamir A., Cohen-Or D.}:
\newblock Non-homogeneous resizing of complex models.
\newblock In \emph{SIGGRAPH Asia} (2006).

\bibitem[LH13]{li2013}
\textsc{Li X.-Y., Hu S.-M.}:
\newblock Poisson coordinates.
\newblock \emph{IEEE Transactions on Visualization and Computer Graphics} (2013).

\bibitem[LLCO08]{green}
\textsc{Lipman Y., Levin D., Cohen-Or D.}:
\newblock Green coordinates.
\newblock In \emph{SIGGRAPH} (2008).

\bibitem[LMR{\etalchar{*}}15]{SMPL:2015}
\textsc{Loper M., Mahmood N., Romero J., Pons-Moll G., Black M.~J.}:
\newblock {SMPL}: A skinned multi-person linear model.
\newblock In \emph{SIGGRAPH Asia} (2015).

\bibitem[LRR{\etalchar{*}}17]{litany2017deep}
\textsc{Litany O., Remez T., Rodola E., Bronstein A., Bronstein M.}:
\newblock {Deep Functional Maps}: Structured prediction for dense shape correspondence.
\newblock In \emph{CVPR} (2017).

\bibitem[LSC{\etalchar{*}}04]{Lipman:2004}
\textsc{{Lipman} Y., {Sorkine} O., {Cohen-Or} D., {Levin} D., {Rossi} C., {Seidel} H.~P.}:
\newblock Differential coordinates for interactive mesh editing.
\newblock In \emph{Shape Modeling Applications} (2004).

\bibitem[LSKK22]{lee2022popoutmotion}
\textsc{Lee J., Sung M., Kim H., Kim T.-K.}:
\newblock {Pop-Out Motion}: {3D}-aware image deformation via learning the shape laplacian.
\newblock In \emph{CVPR} (2022).

\bibitem[LSLCO05]{Lipman:2005}
\textsc{Lipman Y., Sorkine O., Levin D., Cohen-Or D.}:
\newblock Linear rotation-invariant coordinates for meshes.
\newblock In \emph{SIGGRAPH} (2005).

\bibitem[LSMS21]{DeepMetaHandles:2021}
\textsc{Liu M., Sung M., M\v{e}ch R., Su H.}:
\newblock {DeepMetaHandles}: Learning deformation meta-handles of {3D} meshes with biharmonic coordinates.
\newblock In \emph{CVPR} (2021).

\bibitem[LTT{\etalchar{*}}21]{deformthings4d}
\textsc{Li Y., Takehara H., Taketomi T., Zheng B., Niessner M.}:
\newblock {4DComplete}: Non-rigid motion estimation beyond the observable surface.
\newblock In \emph{ICCV} (2021).

\bibitem[MRSHO22]{magnet2022smooth}
\textsc{Magnet R., Ren J., Sorkine-Hornung O., Ovsjanikov M.}:
\newblock Smooth non-rigid shape matching via effective dirichlet energy optimization.
\newblock In \emph{3DV} (2022).

\bibitem[OBCS{\etalchar{*}}12]{functionalmap}
\textsc{Ovsjanikov M., Ben-Chen M., Solomon J., Butscher A., Guibas L.}:
\newblock {Functional Maps}: A flexible representation of maps between shapes.
\newblock \emph{ACM Transactions on Graphics} (2012).

\bibitem[SA07]{Sorkine:2007}
\textsc{Sorkine O., Alexa M.}:
\newblock As-rigid-as-possible surface modeling.
\newblock In \emph{Eurographics Symposium on Geometry Processing} (2007).

\bibitem[Sah20]{shape_corr_survey}
\textsc{Sahillio\u{g}lu Y.}:
\newblock Recent advances in shape correspondence.
\newblock \emph{The Visual Computer} (2020).

\bibitem[SB09]{sorkine09}
\textsc{Sorkine O., Botsch M.}:
\newblock Interactive shape modeling and deformation.
\newblock In \emph{Eurographics Tutorials} (2009).

\bibitem[SCOL{\etalchar{*}}04]{Sorkine:2004}
\textsc{Sorkine O., Cohen-Or D., Lipman Y., Alexa M., R\"{o}ssl C., Seidel H.-P.}:
\newblock Laplacian surface editing.
\newblock In \emph{Eurographics Symposium on Geometry Processing} (2004).

\bibitem[SJA{\etalchar{*}}20]{Sung:2020}
\textsc{Sung M., Jiang Z., Achlioptas P., Mitra N.~J., Guibas L.~J.}:
\newblock {DeformSyncNet}: Deformation transfer via synchronized shape deformation spaces.
\newblock In \emph{SIGGRAPH Asia} (2020).

\bibitem[SP86]{Sederberg:1986}
\textsc{Sederberg T.~W., Parry S.~R.}:
\newblock Free-form deformation of solid geometric models.
\newblock In \emph{SIGGRAPH} (1986).

\bibitem[WBGH11]{webber2011}
\textsc{Weber O., Ben{-}Chen M., Gotsman C., Hormann K.}:
\newblock A complex view of barycentric mappings.
\newblock \emph{Computer Graphics Forum} (2011).

\bibitem[WCMN19]{3dn}
\textsc{Wang W., Ceylan D., Mech R., Neumann U.}:
\newblock {3DN}: {3D} deformation network.
\newblock In \emph{CVPR} (2019).

\bibitem[WJBK15]{Wang:2015:Linear}
\textsc{Wang Y., Jacobson A., Barbic J., Kavan L.}:
\newblock Linear subspace design for real-time shape deformation.
\newblock \emph{ACM Transactions on Graphics} (2015).

\bibitem[WSLG07]{weber2007context}
\textsc{Weber O., Sorkine O., Lipman Y., Gotsman C.}:
\newblock Context-aware skeletal shape deformation.
\newblock \emph{Computer Graphics Forum} (2007).

\bibitem[YAK{\etalchar{*}}20]{Yifan:NeuralCage:2020}
\textsc{Yifan W., Aigerman N., Kim V.~G., Chaudhuri S., Sorkine-Hornung O.}:
\newblock Neural cages for detail-preserving {3D} deformations.
\newblock In \emph{CVPR} (2020).

\bibitem[YM16]{Yumer:2016}
\textsc{Yumer E., Mitra N.~J.}:
\newblock Learning semantic deformation flows with {3D} convolutional networks.
\newblock In \emph{ECCV} (2016).

\bibitem[ZKJB17]{Zuffi:CVPR:2017}
\textsc{Zuffi S., Kanazawa A., Jacobs D., Black M.~J.}:
\newblock {3D Menagerie}: Modeling the {3D} shape and pose of animals.
\newblock In \emph{CVPR} (2017).

\bibitem[ZQZ{\etalchar{*}}21]{zeng2021corrnet3d}
\textsc{Zeng Y., Qian Y., Zhu Z., Hou J., Yuan H., He Y.}:
\newblock {CorrNet3D}: Unsupervised end-to-end learning of dense correspondence for {3D} point clouds.
\newblock In \emph{CVPR} (2021).

\bibitem[ZYDL21]{zheng2021deep}
\textsc{Zheng Z., Yu T., Dai Q., Liu Y.}:
\newblock Deep implicit templates for {3D} shape representation.
\newblock In \emph{CVPR} (2021).

\end{thebibliography}
}

\newif\ifpaper
\papertrue

\setcounter{section}{0}
\renewcommand\thesection{\Alph{section}}
\section{Appendix}
\label{sec:appendix}
\ifpaper
  \newcommand{\refofpaper}[1]{\unskip}
  \newcommand{\refinpaper}[1]{\unskip}
\else
  \makeatletter
  \newcommand{\manuallabel}[2]{\def\@currentlabel{#2}\label{#1}}
  \makeatother
  \manuallabel{sec:intro}{1}
  \manuallabel{sec:related}{2}
  \manuallabel{sec:deformation_handles}{3.1}
  \manuallabel{sec:orig_biharmonic_coor}{3.2}
  \manuallabel{sec:problem_definition}{4.1}
  \manuallabel{sec:continuous_relaxation}{4.2}
  \manuallabel{sec:our_reformulation}{4.3}
  \manuallabel{sec:exp_setup}{5.2}
  \manuallabel{sec:SMPL_SMAL_results}{5.3}
  \manuallabel{sec:Deform4D_results}{5.4}

  \manuallabel{eq:biharmonics_solution}{2}
  \manuallabel{eq:tattas}{6}
  \manuallabel{eq:our_reformulation}{11}

  \manuallabel{fig:teaser}{1}
  \manuallabel{fig:target-driven1}{2}
  \manuallabel{fig:target-driven2}{3}
  
  \manuallabel{tab:target-driven1}{1}
  \manuallabel{tab:semantic}{2}
  \manuallabel{tab:target-driven2}{3}
  
  \newcommand{\refofpaper}[1]{of the main paper}
  \newcommand{\refinpaper}[1]{in the main paper}
\fi

We include more dataset details in this section. Please refer additional qualitative results in the supplementary material.

\subsection{DeformingThings4D~\cite{deformthings4d} Data Used in Our Experiments}
\label{sec:Deform4D_info}
We select seven characters in our experiments namely ~\emph{bear, deer, doggie, dragon moose, procy} and ~\emph{raccoon}, which have more than $1,000$ target shapes. For the per-motion experiments, we use all motions that have more than 100 target shapes for each character, which are namely:
\begin{itemize}
    \item bear3EP: \textbf{Agression}, Drink, Eat1, Eat2, Hide, Idle1, Idle2, Idle3, Lie, Sleep
    \item deerOMG: \textbf{drink}, eat1, hide, Idle1, Idle2, Idle3, lie, sleep
    \item doggieMN5: \textbf{drink}, eat1, eat2, Howl, idle1, idle2, idle3, Lie
    \item dragonOF2: \textbf{act3}, act30, act31, act38, act46, act49, act57
    \item moose1DOG: \textbf{drink}, eat1, Idle1, Idle2, Idle3, lie, Liesleep
    \item procySTEM: \textbf{Actions2}, Idle1, Idle8, Idle9, Idle11, SleepLieSeat0, SleepLieSeat2
    \item raccoonVGG: \textbf{Actions1}, Climb15, Idle0, Idle2, Idle6, SleepLieSeat1
\end{itemize}
The template shape of each category is set to the first frame (bold) of the first animation in the above list.

\onecolumn
\subsection{Additional Qualitative Results of SMPL~\cite{SMPL:2015} and SMAL~\cite{Zuffi:CVPR:2017} with 16 Control Points}
\label{sec:additional_smpl_smal}
\newcommand{\AllComparisonImages}{\Image{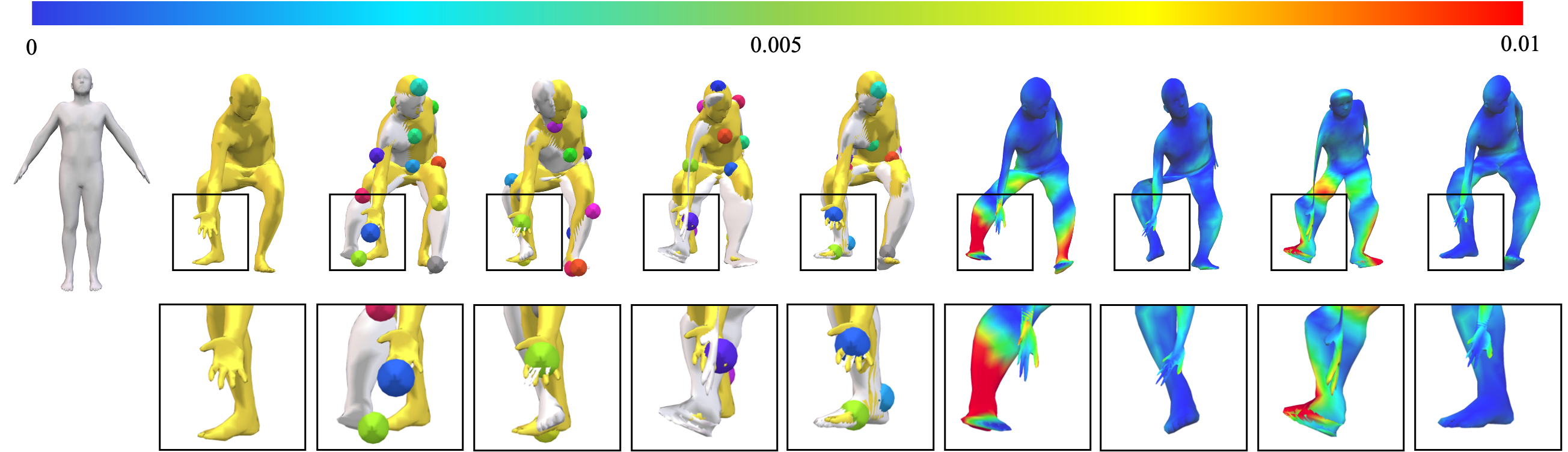}
\Image{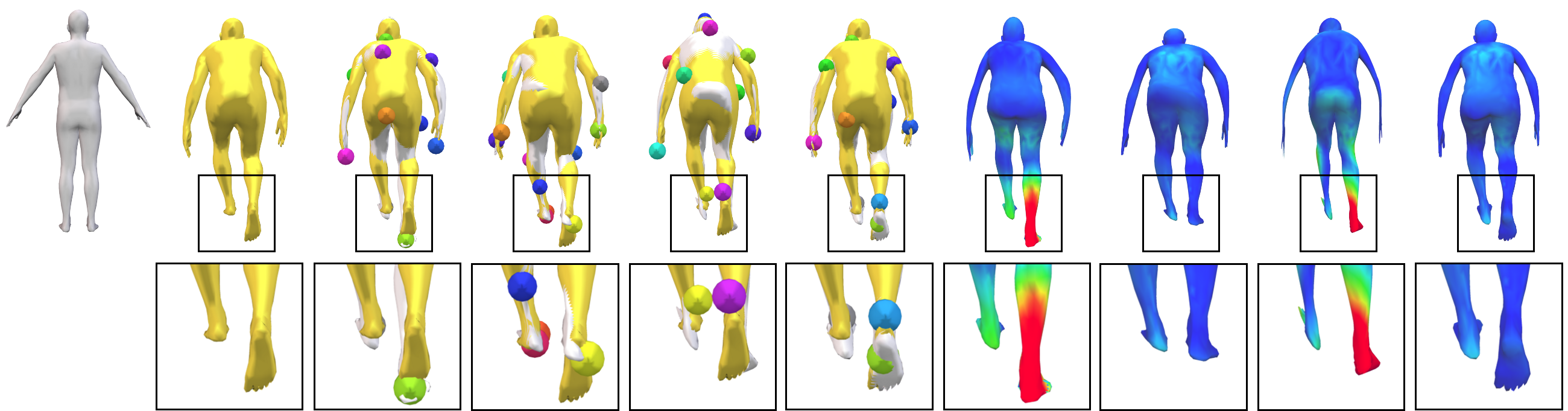}
\Image{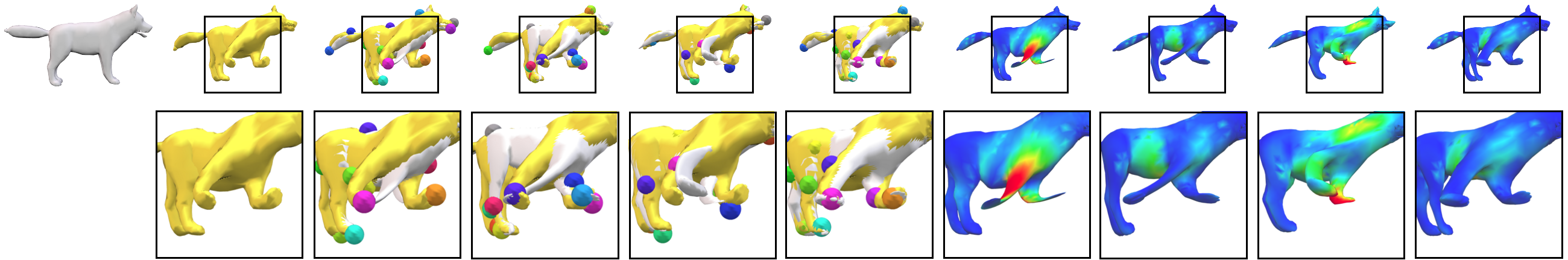}
\Image{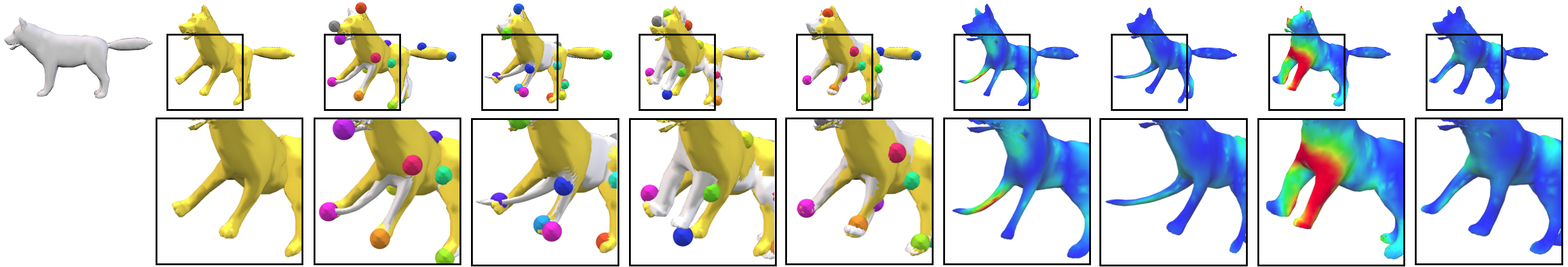}
\Image{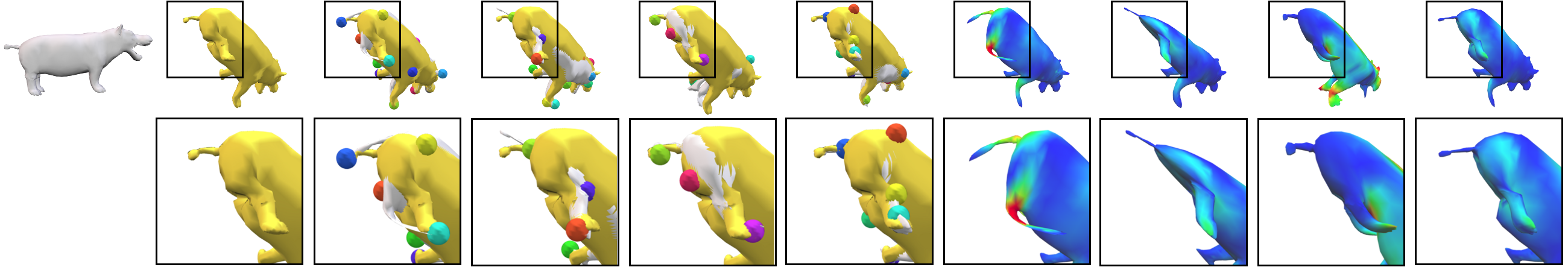}
\Image{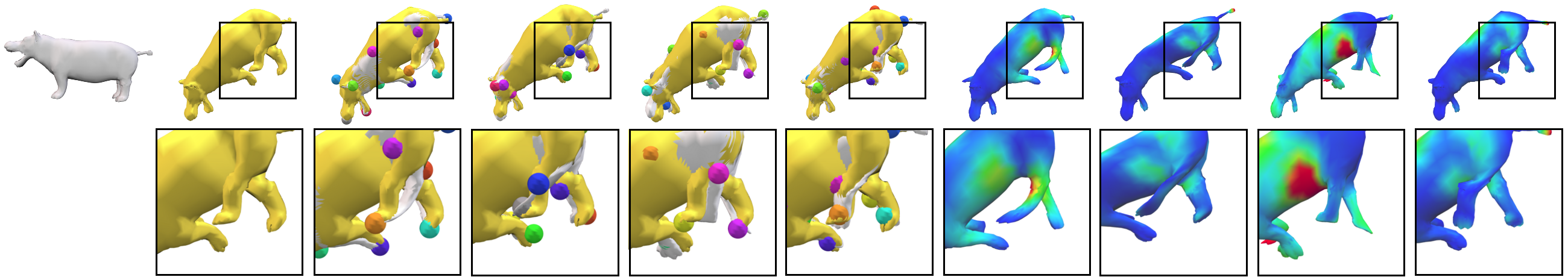}
\Image{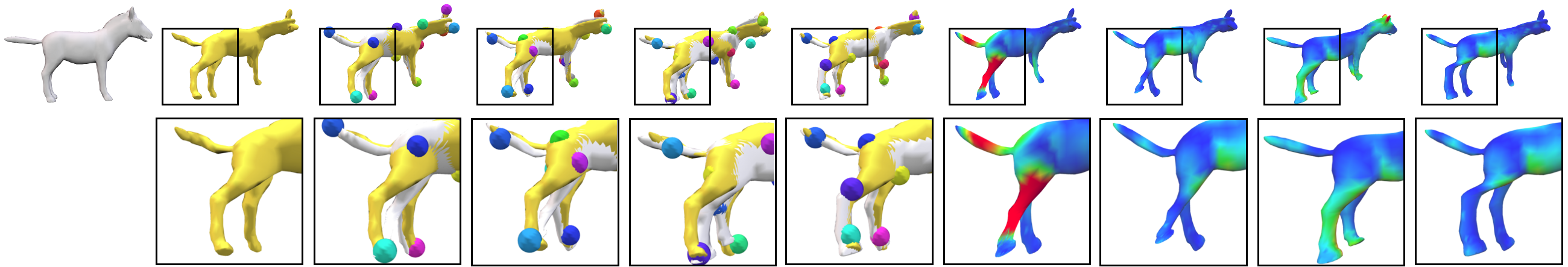}
\Image{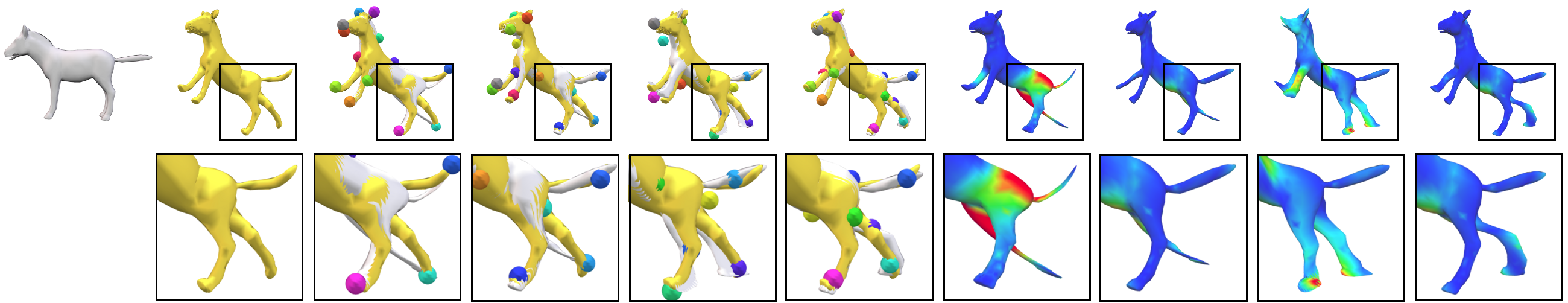}
\Image{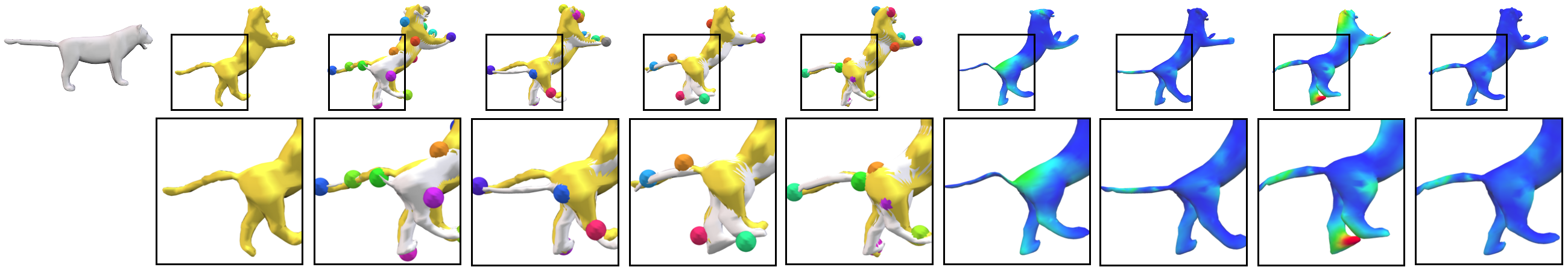}
\Image{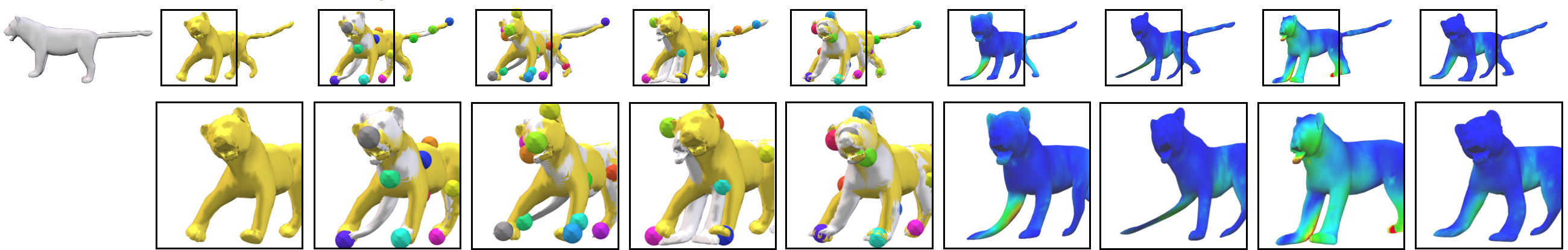}
}
\graphicspath{{figures/fig5/}}

\makeatletter
\def\Image#1{%
  \multicolumn{\LT@cols}{l}{\includegraphics[width=\textwidth]{#1}}\\
}
\makeatother

\LTcapwidth=\textwidth
\setlength{\tabcolsep}{0em}
\def\arraystretch{0.0}
\newcolumntype{Z}{>{\centering\arraybackslash}m{0.1\textwidth}}
{\scriptsize
\begin{longtable}
{ZZ|ZZZZ|ZZZZ}
Template &
Target &
\makecell{FPS}&
\makecell{Random}&
\makecell{KPD}&
\makecell{Ours}&

\makecell{FPS}&
\makecell{Random}&
\makecell{KPD}&
\makecell{Ours} \\
  \midrule
  \endhead

  \bottomrule
  \endfoot

  \Image{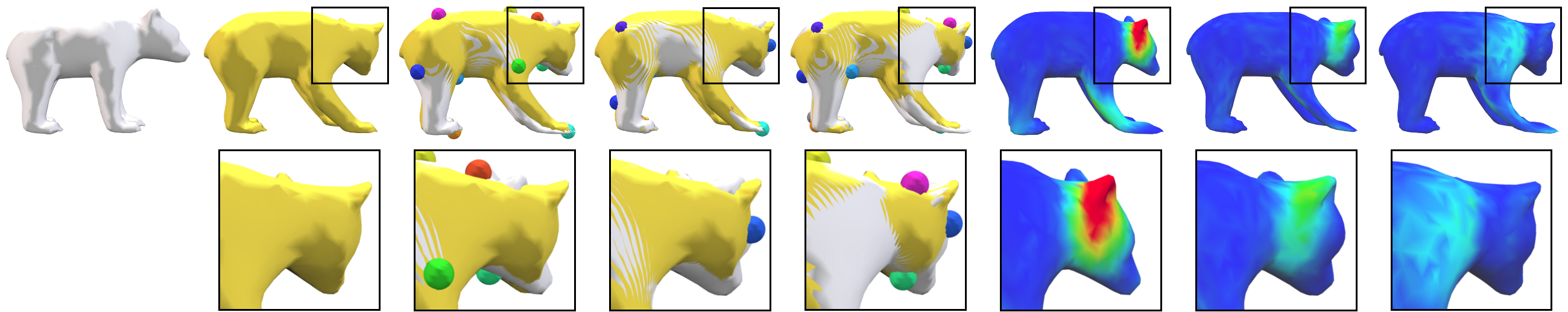}
\Image{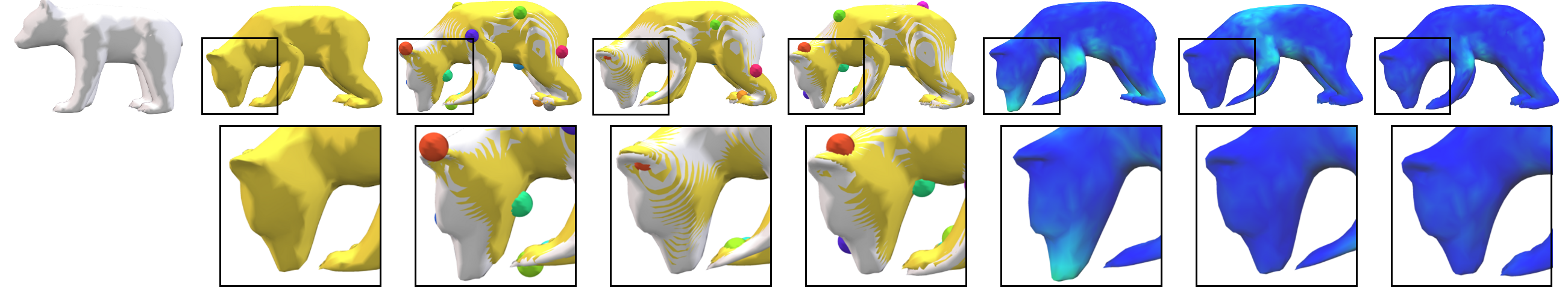}
\Image{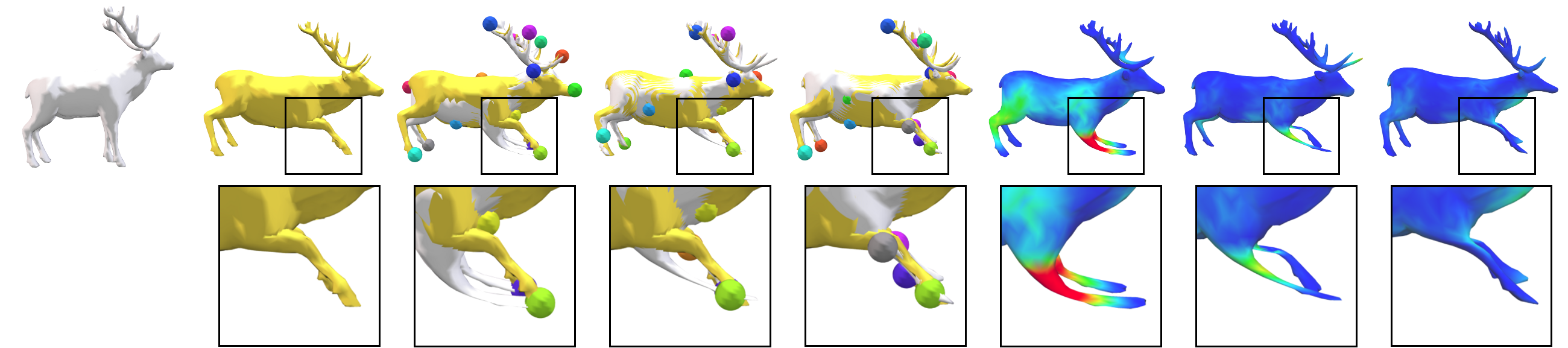}
\Image{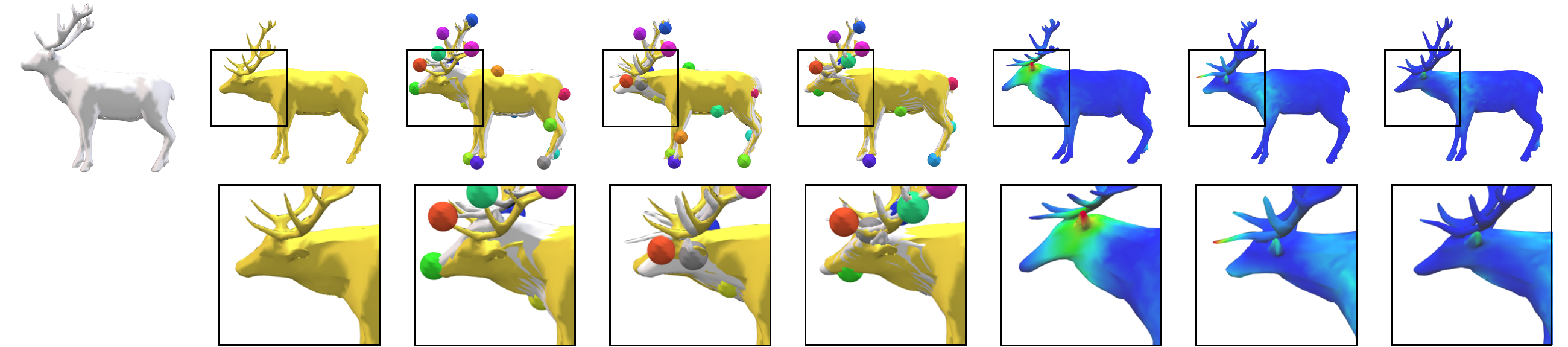}
\Image{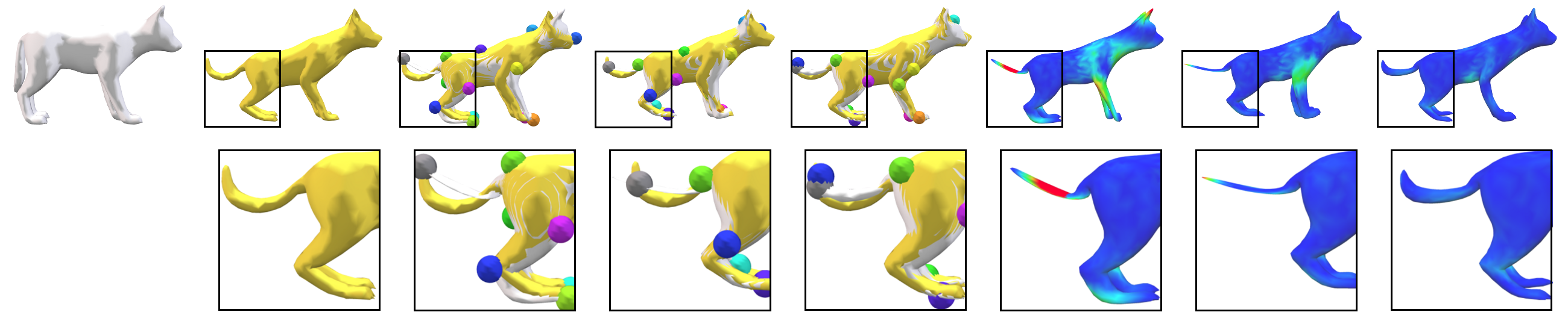}
\Image{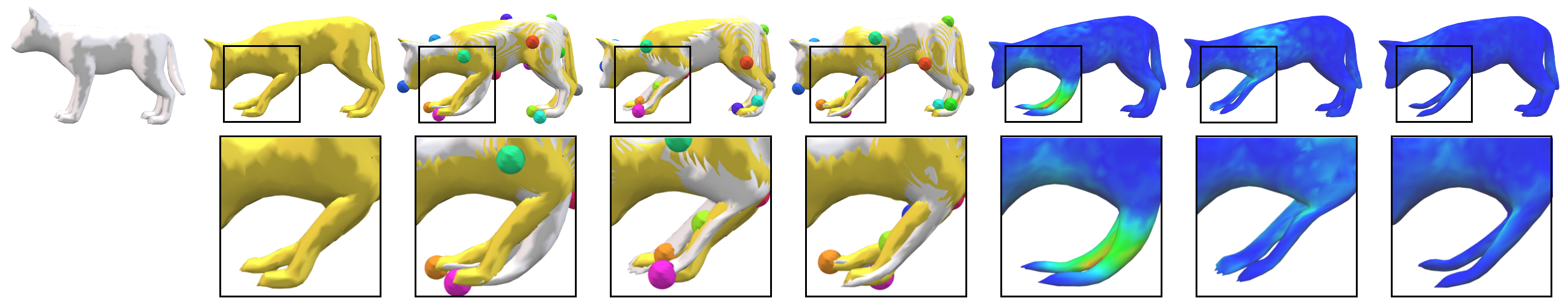}
\Image{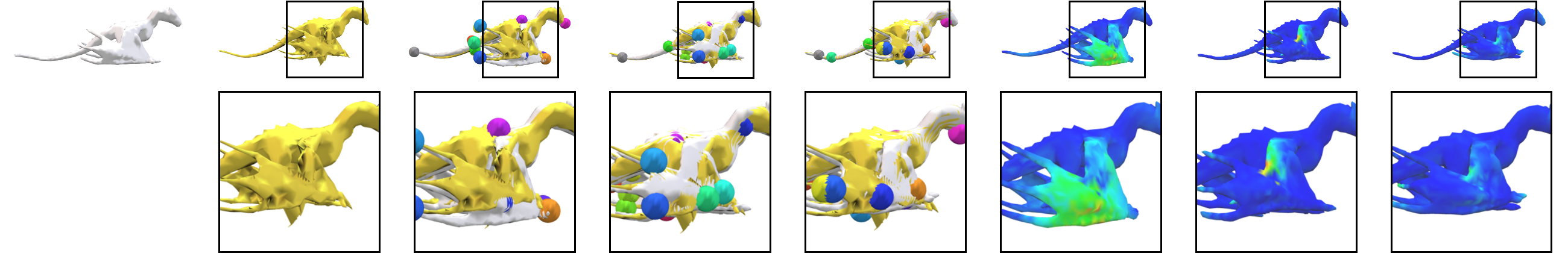}
\Image{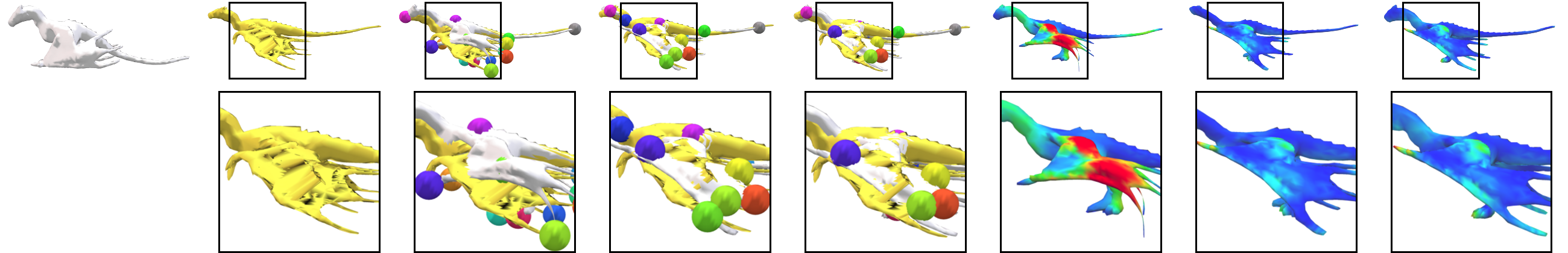}
\Image{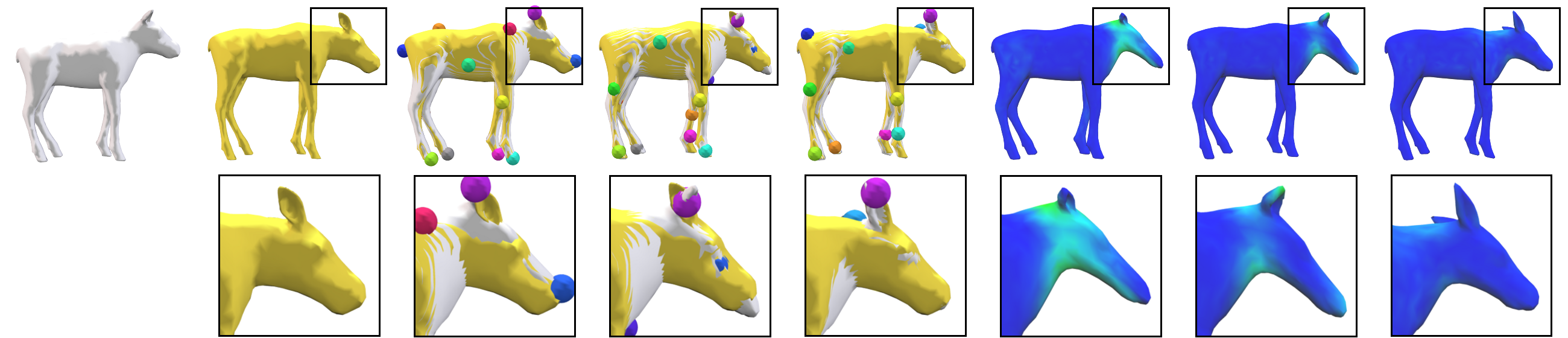}
\Image{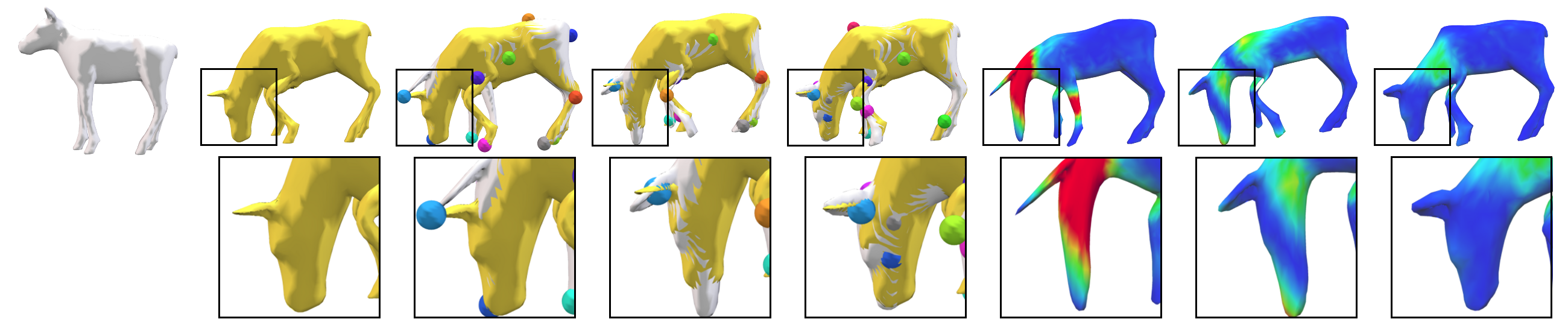}
\Image{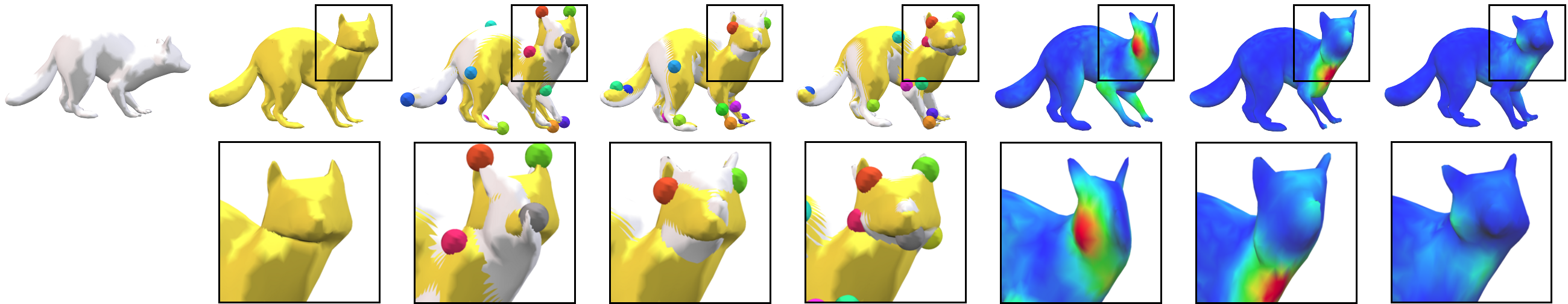}
\Image{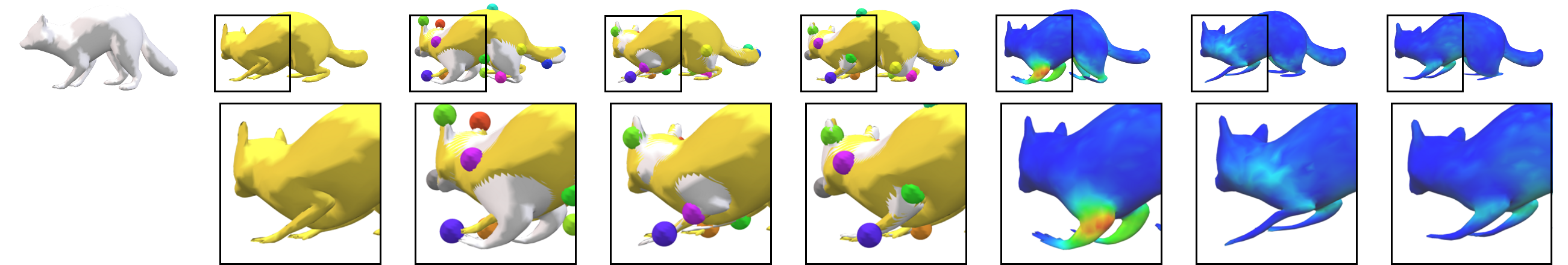}
\Image{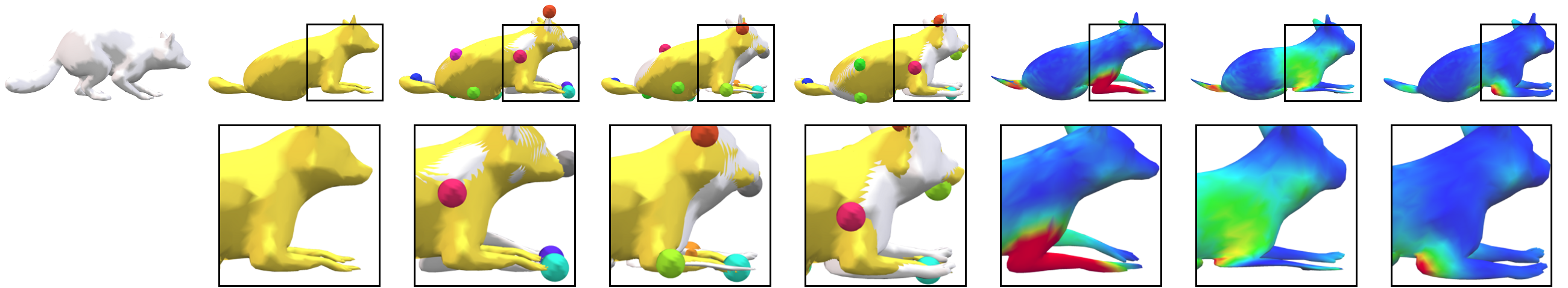}
\Image{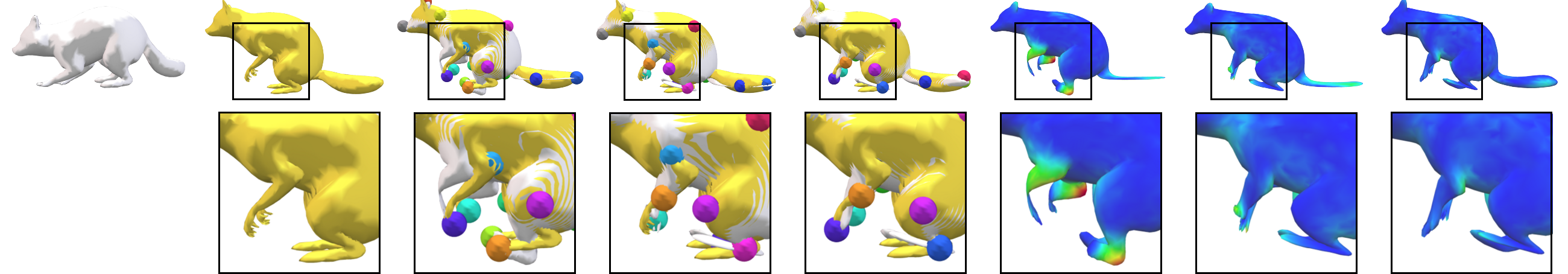}
\end{longtable}
}

\clearpage
\newpage

\subsection{Additional Qualitative Results of SMPL~\cite{SMPL:2015} and SMAL~\cite{Zuffi:CVPR:2017} with 24 Control Points}

\newcommand{\AllComparisonImages}{\Image{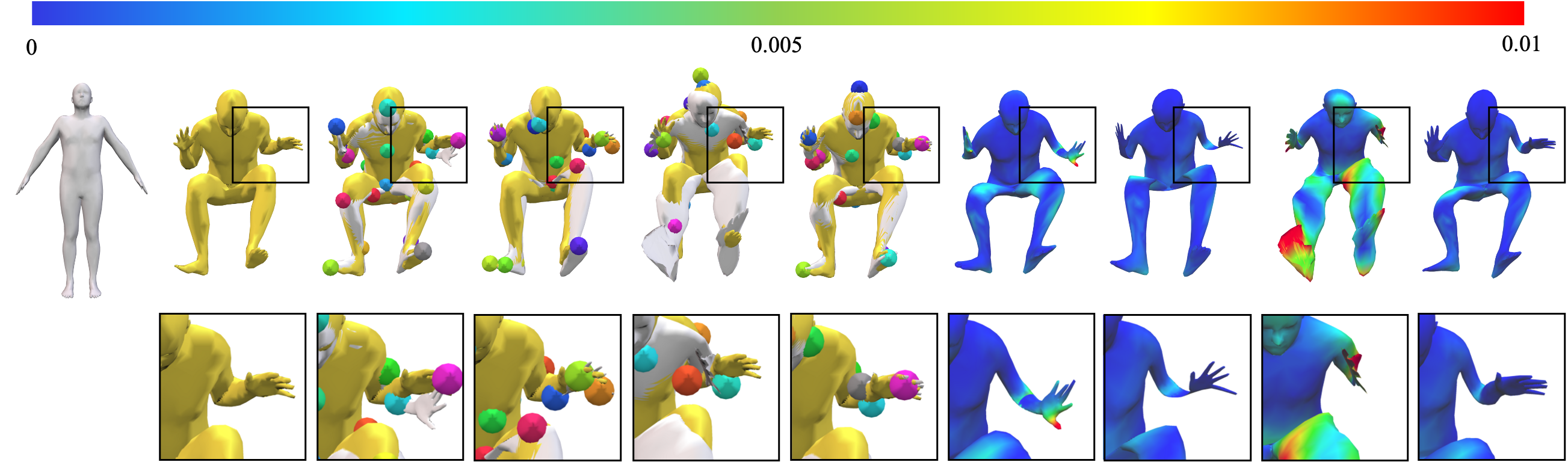}
\Image{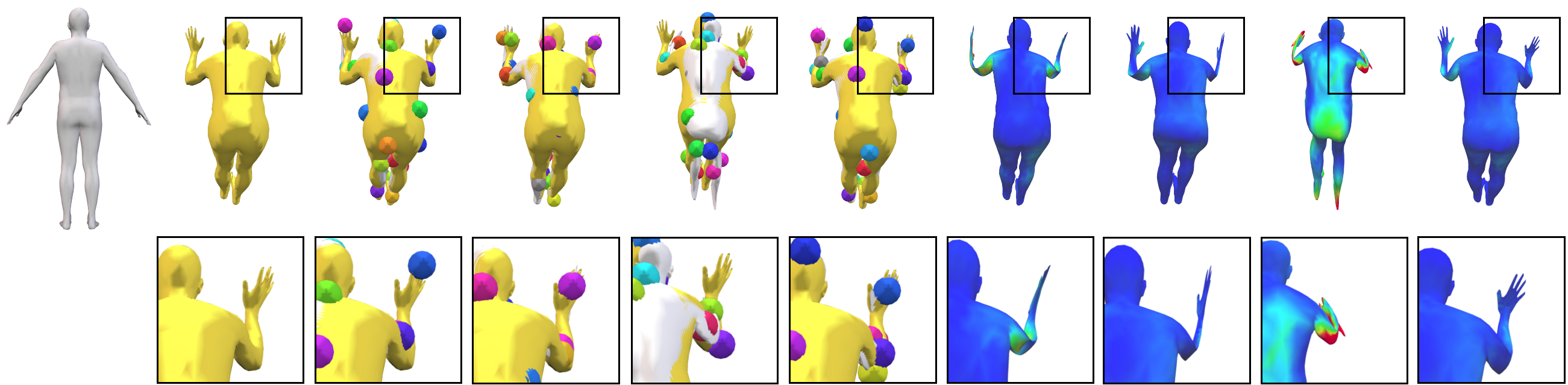}
\Image{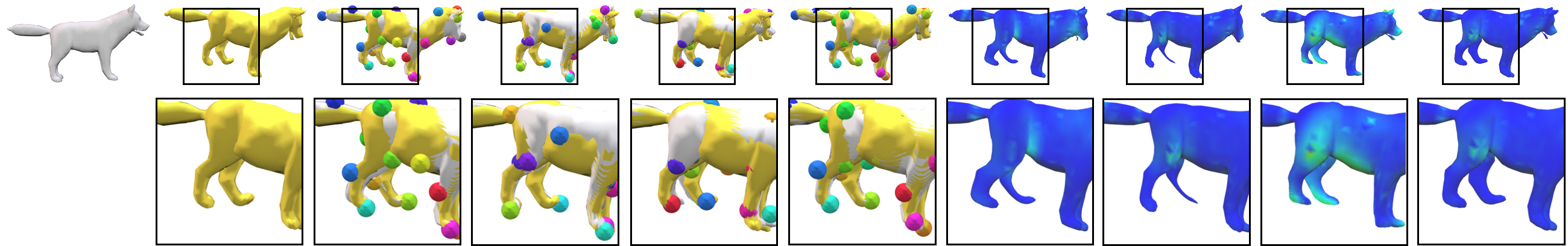}
\Image{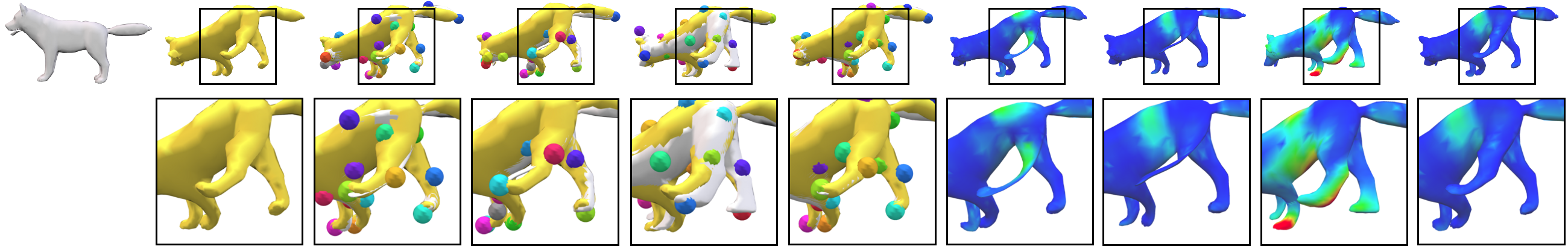}
\Image{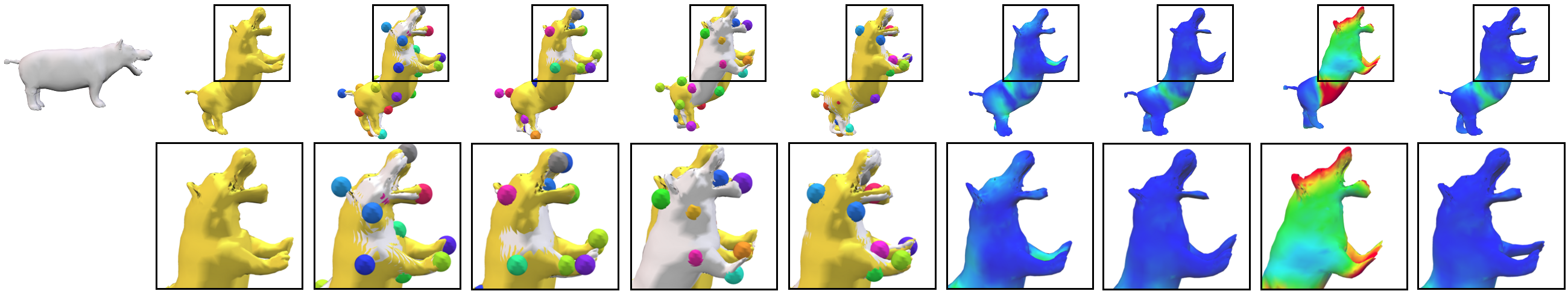}
\Image{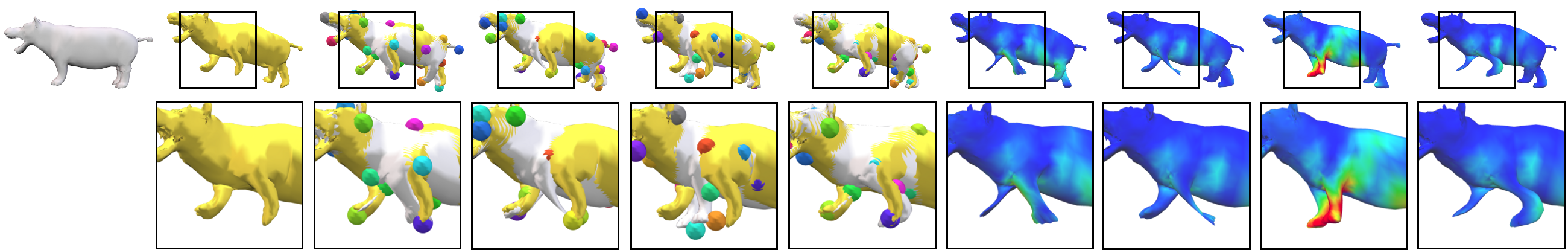}
\Image{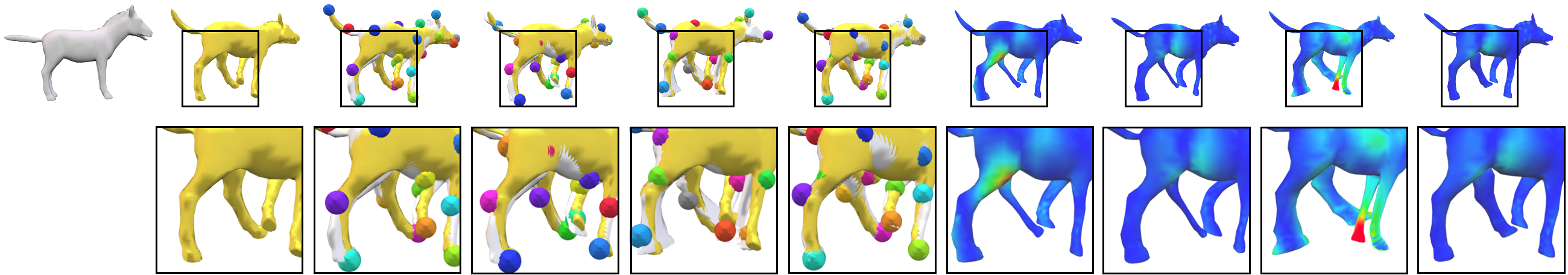}
\Image{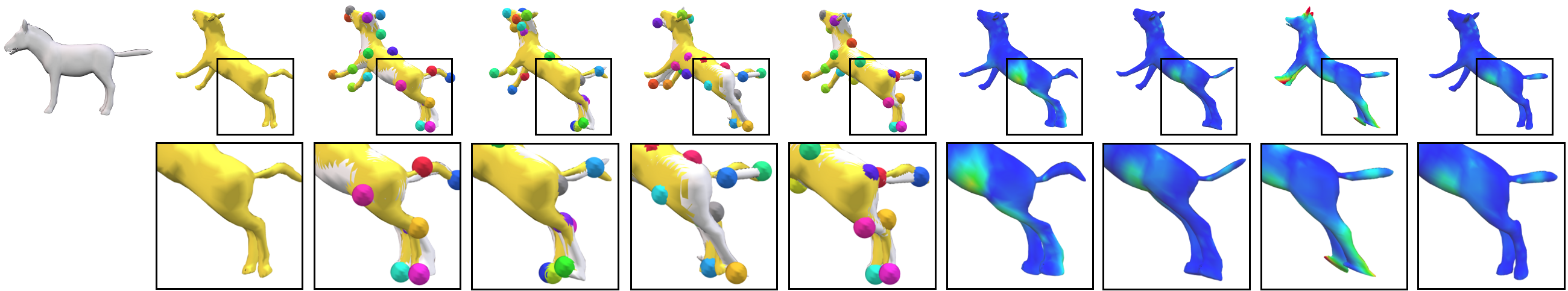}
\Image{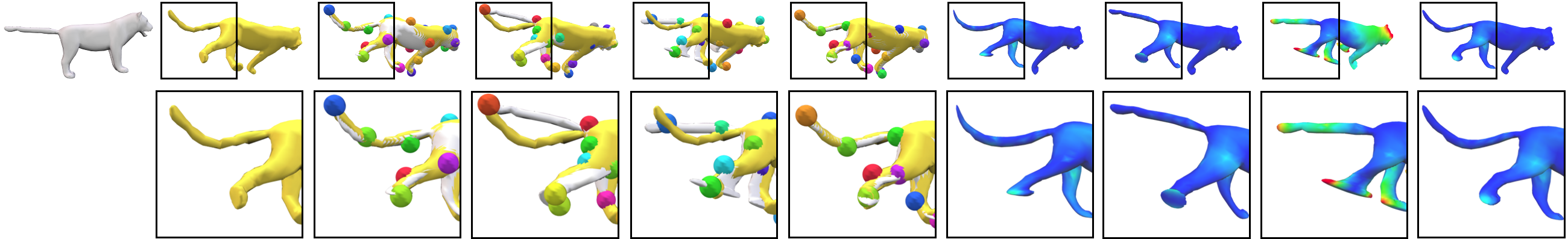}
\Image{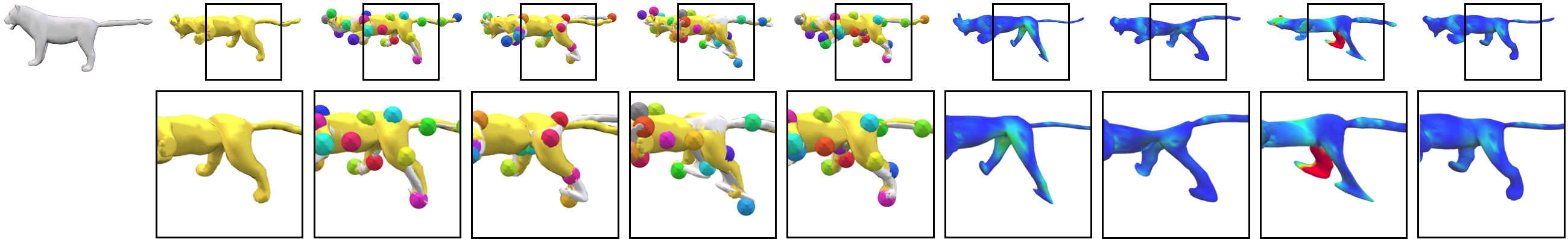}
}
\graphicspath{{figures/fig7/}}

\makeatletter
\def\Image#1{%
  \multicolumn{\LT@cols}{l}{\includegraphics[width=\textwidth]{#1}}\\
}
\makeatother

\LTcapwidth=\textwidth
\setlength{\tabcolsep}{0em}
\def\arraystretch{0.0}
\newcolumntype{Z}{>{\centering\arraybackslash}m{0.1\textwidth}}
{\scriptsize
\begin{longtable}
{ZZ|ZZZZ|ZZZZ}
Template &
Target &
\makecell{FPS}&
\makecell{Random}&
\makecell{KPD}&
\makecell{Ours}&

\makecell{FPS}&
\makecell{Random}&
\makecell{KPD}&
\makecell{Ours} \\
  \midrule
  \endhead

  \bottomrule
  \endfoot

  \Image{figures/fig6/00.png}
\Image{figures/fig6/01.png}
\Image{figures/fig6/02.png}
\Image{figures/fig6/03.png}
\Image{figures/fig6/04.png}
\Image{figures/fig6/05.png}
\Image{figures/fig6/06.png}
\Image{figures/fig6/07.png}
\Image{figures/fig6/08.png}
\Image{figures/fig6/09.png}
\Image{figures/fig6/10.png}
\Image{figures/fig6/11.png}
\Image{figures/fig6/12.png}
\Image{figures/fig6/13.png}
\end{longtable}
}

\clearpage
\newpage

\subsection{Additional Qualitative Results of SMPL~\cite{SMPL:2015} and SMAL~\cite{Zuffi:CVPR:2017} with 32 Control Points}

\newcommand{\AllComparisonImages}{\Image{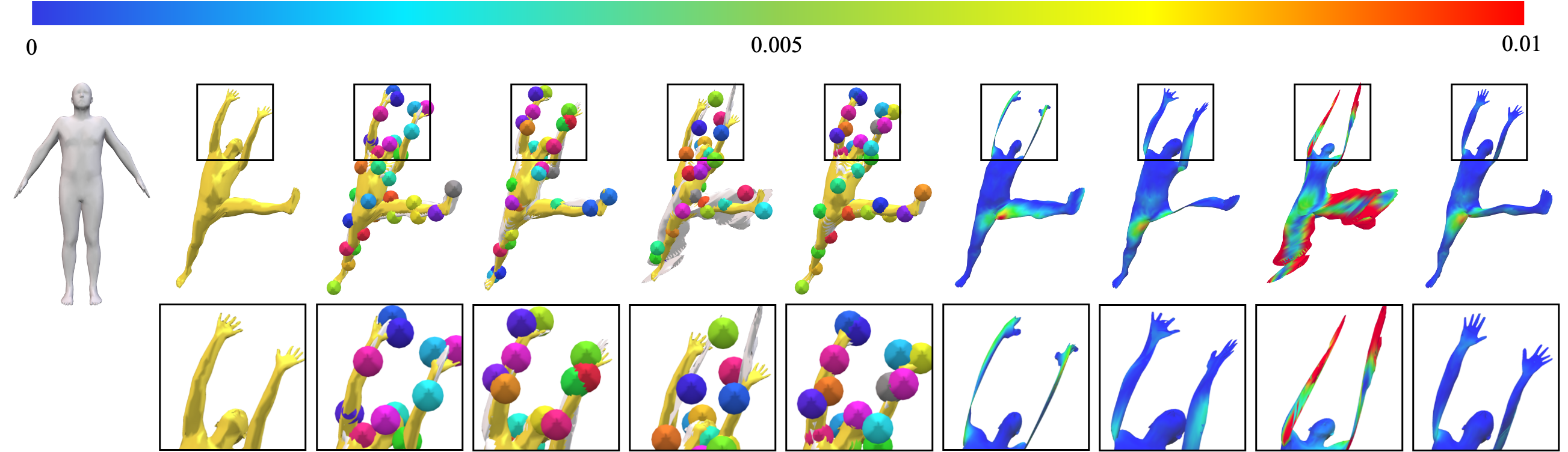}
\Image{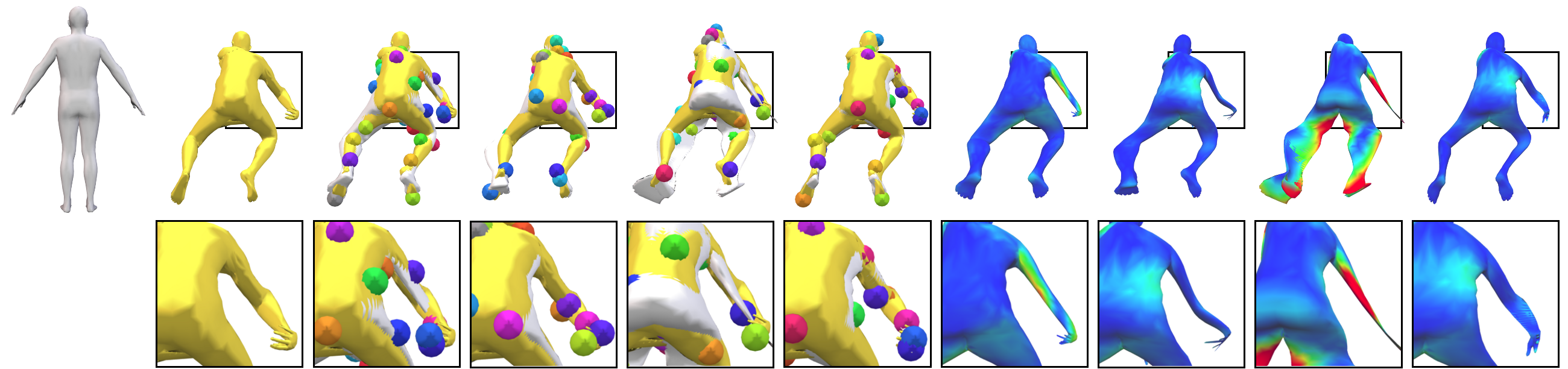}
\Image{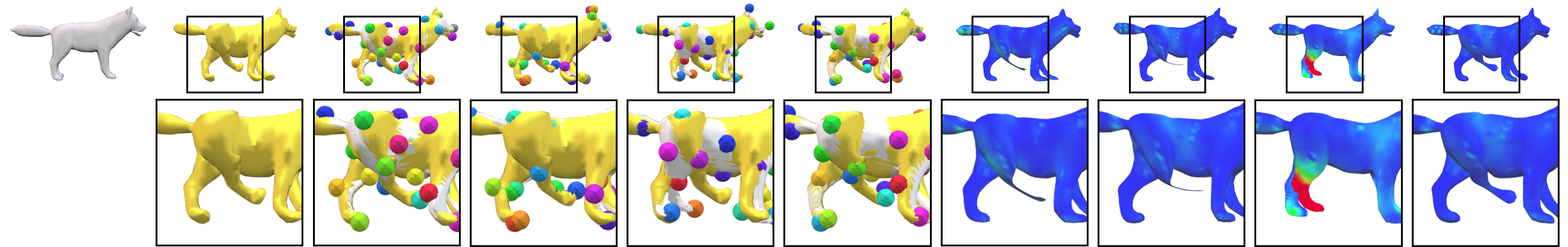}
\Image{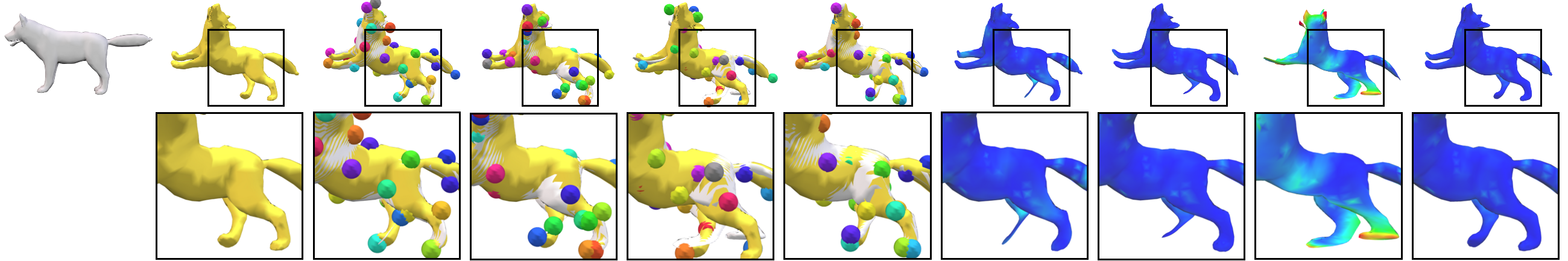}
\Image{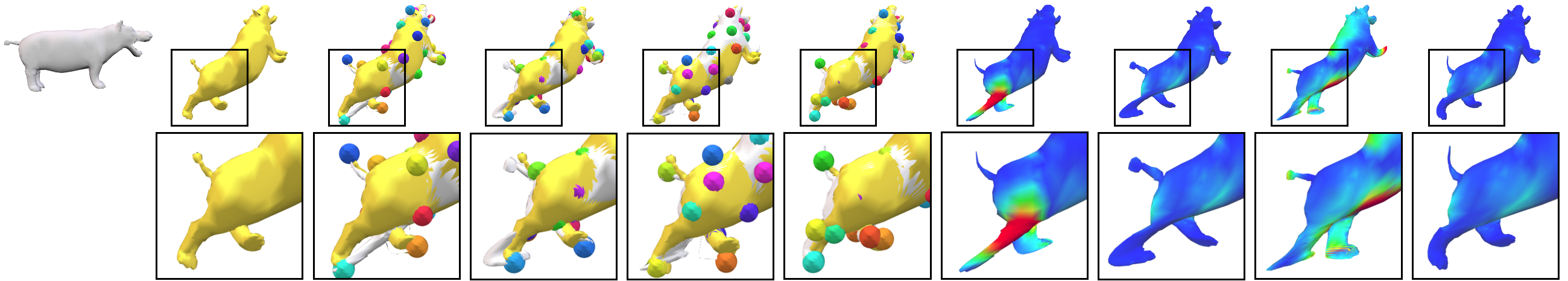}
\Image{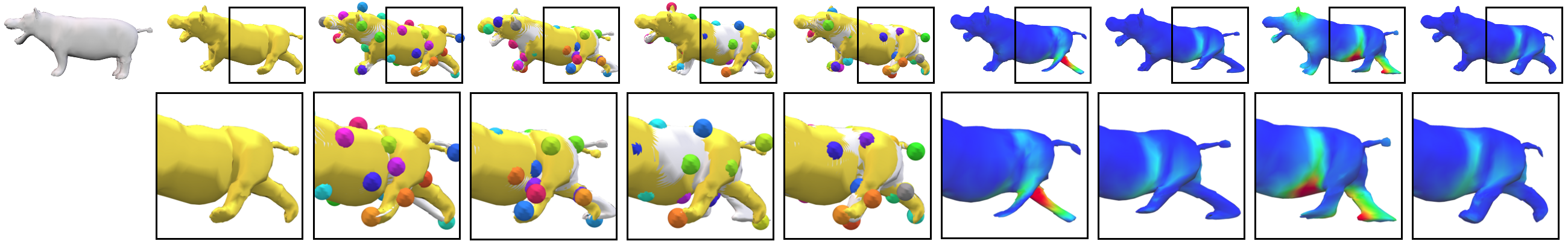}
\Image{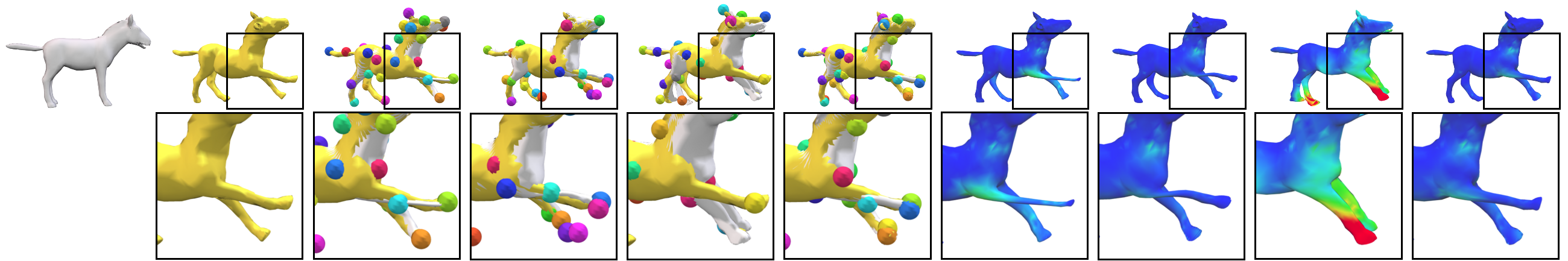}
\Image{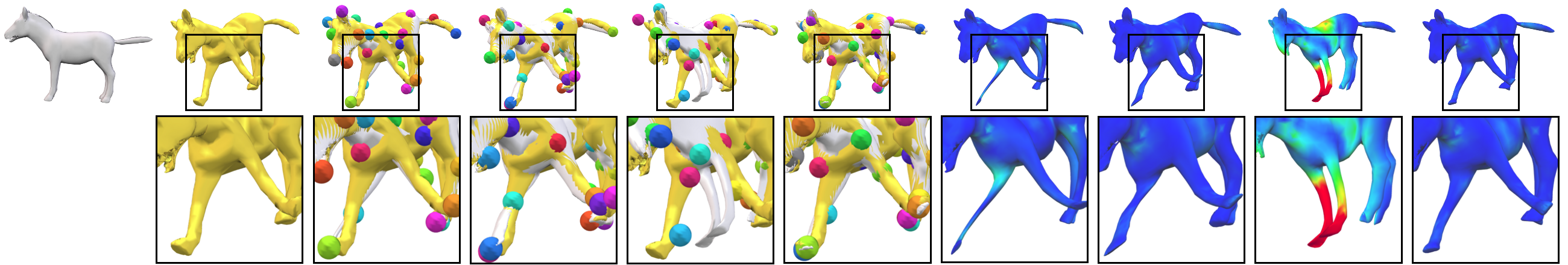}
\Image{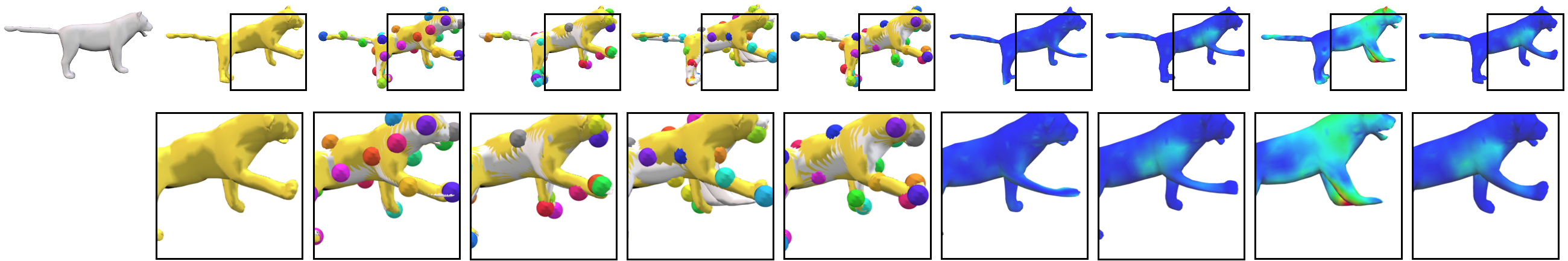}
\Image{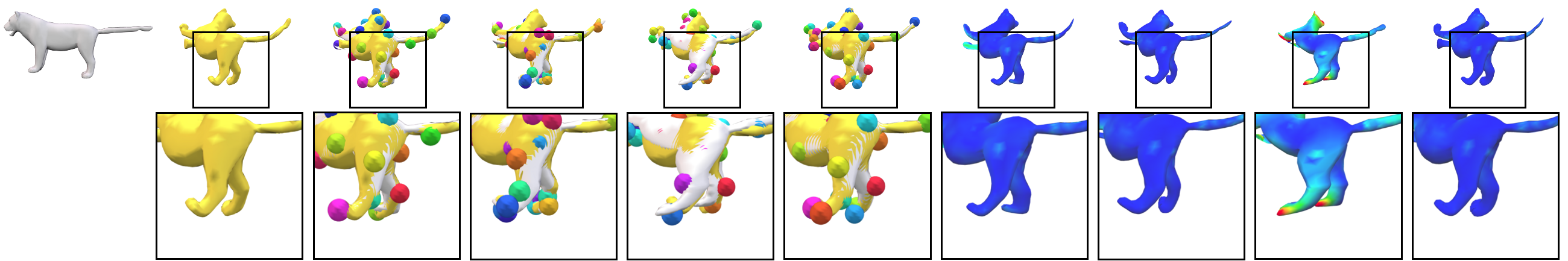}}
\graphicspath{{figures/fig8/}}

\makeatletter
\def\Image#1{%
  \multicolumn{\LT@cols}{l}{\includegraphics[width=\textwidth]{#1}}\\
}
\makeatother

\LTcapwidth=\textwidth
\setlength{\tabcolsep}{0em}
\def\arraystretch{0.0}
\newcolumntype{Z}{>{\centering\arraybackslash}m{0.1\textwidth}}
{\scriptsize
\begin{longtable}
{ZZ|ZZZZ|ZZZZ}
Template &
Target &
\makecell{FPS}&
\makecell{Random}&
\makecell{KPD}&
\makecell{Ours}&

\makecell{FPS}&
\makecell{Random}&
\makecell{KPD}&
\makecell{Ours} \\
  \midrule
  \endhead

  \bottomrule
  \endfoot

  \Image{figures/fig6/00.png}
\Image{figures/fig6/01.png}
\Image{figures/fig6/02.png}
\Image{figures/fig6/03.png}
\Image{figures/fig6/04.png}
\Image{figures/fig6/05.png}
\Image{figures/fig6/06.png}
\Image{figures/fig6/07.png}
\Image{figures/fig6/08.png}
\Image{figures/fig6/09.png}
\Image{figures/fig6/10.png}
\Image{figures/fig6/11.png}
\Image{figures/fig6/12.png}
\Image{figures/fig6/13.png}
\end{longtable}
}

\clearpage
\newpage

\subsection{Additional Qualitative Results with DeformingThings4D~\cite{deformthings4d} with 16 Control Points}
\label{sec:additional_deform4d}
\newcommand{\AllComparisonImages}{\Image{figures/fig6/00.png}
\Image{figures/fig6/01.png}
\Image{figures/fig6/02.png}
\Image{figures/fig6/03.png}
\Image{figures/fig6/04.png}
\Image{figures/fig6/05.png}
\Image{figures/fig6/06.png}
\Image{figures/fig6/07.png}
\Image{figures/fig6/08.png}
\Image{figures/fig6/09.png}
\Image{figures/fig6/10.png}
\Image{figures/fig6/11.png}
\Image{figures/fig6/12.png}
\Image{figures/fig6/13.png}}
\graphicspath{{figures/fig6/}}

\makeatletter
\def\Image#1{%
  \multicolumn{\LT@cols}{l}{\includegraphics[width=\textwidth]{#1}}\\
}
\makeatother

\LTcapwidth=\textwidth
\setlength{\tabcolsep}{0em}
\def\arraystretch{0.0}
\newcolumntype{Z}{>{\centering\arraybackslash}m{0.125\textwidth}}
{\scriptsize
\begin{longtable}
{ZZ|ZZZ|ZZZ}
Template &
Target &
\makecell{FPS}&
\makecell{Ours Per-category}&
\makecell{Ours Per-motion}&

\makecell{FPS}&
\makecell{Ours Per-category}&
\makecell{Ours Per-motion}\\
  \midrule
  \endhead

  \bottomrule
  \endfoot

  \Image{figures/fig6/00.png}
\Image{figures/fig6/01.png}
\Image{figures/fig6/02.png}
\Image{figures/fig6/03.png}
\Image{figures/fig6/04.png}
\Image{figures/fig6/05.png}
\Image{figures/fig6/06.png}
\Image{figures/fig6/07.png}
\Image{figures/fig6/08.png}
\Image{figures/fig6/09.png}
\Image{figures/fig6/10.png}
\Image{figures/fig6/11.png}
\Image{figures/fig6/12.png}
\Image{figures/fig6/13.png}
\end{longtable}
}

\end{document}